\documentclass{vldb}
\usepackage{graphicx}
\usepackage{algorithm}
\usepackage{algorithmicx}
\usepackage{algpseudocode}
\usepackage{balance}  
\usepackage{color}

\usepackage{url}

\usepackage{subfig}
\usepackage{tabularx}
\algnewcommand{\LineComment}[1]{\State \(\triangleright\) #1}

\begin{document}



\title{BigSparse: High-performance external graph analytics}



%
%
%
%

\numberofauthors{1} 

\author{
%
%
\alignauthor
Sang-Woo Jun$^\dagger$, Andy Wright$^*$, Sizhuo Zhang$^*$, Shuotao Xu$^\dagger$ and Arvind$^\dagger$\\
       \affaddr{Department of Electrical Engineering and Computer Science}\\
       \affaddr{Massachusetts Institute of Technology}\\
	   \email{$^\dagger$\{wjun,shuotao,arvind\}@csail.mit.edu}
	   \email{$^*$\{acwright,szzhang\}@mit.edu}
}

\maketitle

\begin{abstract}

We present BigSparse, a fully external graph analytics system that picks up where semi-external systems like FlashGraph and X-Stream, which only store vertex data in memory, left off. 
BigSparse stores both edge and vertex data in an array of SSDs and avoids random updates to the vertex data, by first logging the vertex updates and then sorting the log to sequentialize accesses to the SSDs.
This newly introduced sorting overhead is reduced significantly by interleaving sorting with vertex reduction operations.
In our experiments on a server with 32GB to 64GB of DRAM, 
BigSparse outperforms other in-memory and semi-external graph analytics systems for algorithms such as PageRank, Breadth-First Search, and Betweenness-Centrality for terabyte-size graphs with billions of vertices. 
BigSparse is capable of high-speed analytics of much larger graphs, on the same machine configuration.

\end{abstract}


\section{Introduction}

Extremely large and sparse graphs with irregular structures (billions of vertices and hundreds of billions of edges) arise in many important problems, for example, analyses of social networks, causal modeling in large multi-sensor data sets, and high dimensional feature networks. 
Their efficient processing enables everything from optimal resource management in power grids~\cite{smartgrid} to terrorist network detection~\cite{terrorists}.

Due to the low locality and data-intensive nature of graph analytics~\cite{challengesparallelgraph}, caching becomes ineffective, and as a result, performance becomes bound to the performance of random-accesses to data storage.
Since small-granularity random-accesses in secondary storage are much more expensive than in DRAM, traditional systems for graph computation are designed with enough DRAM to hold the entire graph; it is simply accepted that the system doesn't function when the data does not fit the available DRAM.
This approach can be costly; a scale 36 Graph 500 benchmark graph has $2^{36}$ vertices and requires approximately 17 TBs of storage~\cite{graph500}.
We would need 120 servers with 128 GBs of memory each to process a graph of this magnitude.
Such a system would cost well over \$300K and consume over 20 KW of electricity.
Besides the capital cost, redesigning a single-node system to run on a distributed and scalable system introduces its own inefficiencies, and additional performance limitations imposed by interconnection networks.
As a result, it often takes many nodes in a multi-node system to scale past an optimized single-threaded application~\cite{mcsherry2015scalability}.

An alternative approach is to use secondary storage which is much cheaper and cooler than DRAM; 
even a single node can accommodate 1TB to 16TBs of data with current NAND flash technology. 
Processing in secondary storage, however, presents a different set of challenges than processing in DRAM: 
(1) bandwidth is much lower; 
(2) latency is orders of magnitude higher; and 
(3) access granularity is 4KB to 8KB, as opposed to a cache line.
Large access granularity can create a serious storage bandwidth problem for fine-grained random access; if only 4 bytes of data from an 8KB flash page is used, the bandwidth is reduced effectively by a factor of 2000. 
Hard disks require even larger granularity sequential accesses to overcome the seek-time penalty. 
As a result, naive use of secondary storage as a memory extension will result in a sharp performance drop, to the point where the system is no longer viable.

There have been several attempts to overcome the limitations of secondary storage with different degrees of success.
One of the earliest systems, GraphChi~\cite{graphchi}, is a fully external system that stores all the data in external storage and consequently, requires extremely small amounts of DRAM. 
It re-organizes the algorithm to make data access completely sequential, and thus, make accesses perfectly suitable for coarse-grained disk access. 
However, GraphChi does so by introducing additional work, and requires the whole graph data to be read multiple times each iteration. These extra calculations result in low performance on large graphs and makes GraphChi uncompetitive with memory-based systems.

On the other hand, FlashGraph~\cite{flashgraph} and X-Stream~\cite{xstream} follow a \emph{semi-external} approach by storing only the edge data in external storage while keeping the much smaller vertex data in DRAM. 
These systems restructure the algorithm in such a manner that the edge data is accessed sequentially or in coarse granularities, and random accesses are restricted to the vertex data in DRAM.
Such systems require much less DRAM than in-memory systems, and yet achieve performance that is competitive with in-memory systems.
However, they do have some limitations. X-Stream iterates through the entire graph every iteration, making it less effective with sparse algorithms. 
FlashGraph demonstrates performance rivalling in-memory systems, but requires the entire vertex data to fit in memory, otherwise the performance degrades quickly.

In this paper we present BigSparse, a novel \emph{fully external} graph analytics system that achieves high performance while drastically reducing the memory requirements.
BigSparse achieves competitive performance compared to in-memory and semi-external graph analytics systems while requiring constant memory space regardless of problem size, and can provide a graceful degradation in performance as the vertex file size exceeds the memory capacity.
BigSparse has an order of magnitude lower memory requirement compared to semi-external systems like FlashGraph~\cite{flashgraph}, and multiple orders of magnitude lower memory requirement compared to in-memory systems such as GraphLab and PowerGraph~\cite{graphlab, powergraph}, 
In order to overcome the fine-grained random access overhead, BigSparse first logs all the vertex updates in an external list, and then sorts and reduces this list before updating the vertex data.
The overhead of sorting the update log is further reduced by interleaving sorting and reduction operations. This interleaving greatly reduces the log size at every sort phase and consequently, the overhead of sorting is more than paid for by the advantage of sequential storage accesses.

We have evaluated the performance of a single-node BigSparse for applications including breadth-first-search, betweenness-centrality and PageRank on the Web Data Commons graph~\cite{wdcgraph} with 3.5 billion vertices, and a suite of synthetic graphs ranging from 10GB to 200GB in size.
We have compared BigSparse against various in-memory, semi-external and external systems under many different system configurations, and show that as graph sizes become larger relative to memory capacity, BigSparse is able to outperform all other systems.
For example, given a modest system with 48GBs of memory to analyze the Web Domain Commons graph, BigSparse was the only system that could still perform high-speed analytics in all algorithms tested.
Other systems either started thrashing swap space because of insufficient memory, or incurred too much additional computation.
Semi-external systems performed well when provided with large enough memory to process large graphs.
For example, given sufficient memory of 96GB or more to perform PageRank on the web graph, BigSparse performs on par with X-Stream, and FlashGraph performs faster than BigSparse.
However, graphs of interest will continue to grow.
In fact, it is likely that the size of available graph datasets are limited by the performance of analytics platforms on affordable machines.
As graph sizes become larger, fast and fully external systems like BigSparse are an attractive way to achieve affordable, high performance graph analytics.

The major contributions of this paper are as follows:
\begin{itemize}
\setlength\itemsep{.1em}
\item The novel Sort-Reduce method of vertex updates to achieve high performance graph analytics on secondary storage, including the important optimization of interleaving sorting and reduction operations.
\item The measured performance of BigSparse using multiple applications under a variety of system configurations, and show that it performs as well as in-memory and semi-external systems while requiring much less memory.
\item An in-depth analysis of the performance impact of our innovations on a variety of system configurations
\end{itemize}

The rest of the paper is organized as follows:
Section~\ref{sec:related} introduces existing research in large scale graph analytics, and Section~\ref{sec:sortreduce} describes our novel Sort-Reduce method of sequentializing random access in graph analytics.
Section~\ref{sec:design} describes the implementation of BigSparse in detail, and its performance evaluation is presented in Section~\ref{sec:evaluation}. We conclude in Section~\ref{sec:conclusion} and provide future research directions.

\section{Related Work}
\label{sec:related}
\subsection{Large Graph Analytics}

Large-scale graph analytics are usually done using a distributed graph processing platform so the user does not have to deal with the difficulties of distribution and parallelism.
Instead the user adapts their graph algorithm to the programming model provided by the graph processing platform.
Not only do these platforms differ in their use of DRAM and secondary storage, they also differ in their programming models and execution models.

Many prominent graph analytics platforms, including Pregel~\cite{pregel} and GraphLab~\cite{graphlab}, expose a vertex-centric programming
model because of its ease of distributed execution. 
In a vertex-centric model, a graph algorithm is deconstructed so that it can be represented by running a \emph{vertex program} on each of the vertices.
A vertex program takes as input information about the current vertex, as well as its neighboring vertices and the edges that connect them. 
After execution, a vertex program updates the current vertex, and possibly sends messages to neighboring vertices.
Vertex-centric systems can further be categorized into two paradigms; Pull-style systems, in which the program reads the values of neighboring vertices and updates its own value, and Push-style systems, in which the program updates the values of its neighbors.
In the push-style system, each vertex's value is updated as many times as its incoming neighbors, whereas in pull-style systems updates happen once for each vertex.

On the other hand, some systems such as X-Stream~\cite{xstream} provide edge-centric programming models, which is aimed at sequentializing accesses into edge data stored in secondary storage, which is usually significantly larger than vertex data.
X-Stream has the benefits of doing completely sequential accesses to the edge data, but has some limitations, such as requiring the processing of all edges in the graph at every cycle, rendering it inefficient for algorithms with sparse active lists like breadth-first search.
X-Stream logs updates to vertex values before applying them, so when available memory capacity is low they can easily partition operations without losing much performance, by simply splitting the stream.
This kind of partitioning is not readily applicable to vertex-centric systems such as FlashGraph~\cite{flashgraph}, because reading the value of neighboring vertices requires fine-grained random access.
Some systems, including Ligra~\cite{ligra}, aim to optimize performance for data that can fit in the memory of a single machine, by making the best use of shared memory parallel processing.

Some recent systems such as Combinatorial BLAS~\cite{buluc2011combinatorial}, GraphBLAS~\cite{graphblas}, and Graphulo~\cite{graphulo} provide the user a set of linear algebra functions designed for graph algorithms~\cite{kepner2011graph}.
These functions are very flexible since they abstract over a user-defined \emph{semiring}.
A semiring is an algebraic structure consisting of two operators, Addition and Multiplication.
By multiplying the graph's adjacency matrix with a vector of vertex values over a custom semiring, the user can specify how the edge weights in the matrix and the vertex values in the vector are \emph{multiplied}, and how the partial results are \emph{added} together to implement the user's desired graph algorithm.
This matrix-vector multiplication can be viewed as a synchronous vertex-centric model restricted to semiring operations.

While both the vertex-centric and linear-algebraic models are versatile enough for many important applications, some applications still require finer grained control of execution. 
Galois~\cite{galois-graph} is one such framework and it has high-performance platform implementations both on shared and distributed memory systems.

Not only do these platforms differ on how the algorithm is expressed, they also differ on how the algorithms are executed.
Systems such as Pregel and the linear-algebraic platforms follow a Bulk Synchronous Parallel (BSP) model of execution, where computation is organized into \emph{supersteps}.
A \emph{superstep} can only see values from the previous superstep, and
supersteps are separated by global barriers, so that each superstep happens in lock-step. 
A BSP based system is easy to parallelize, but incurs the overhead of synchronization. On the other hand, systems such as GraphLab and PowerGraph~\cite{graphlab, powergraph} follow an asynchronous model, in which results of each computation is immediately visible to other vertices. It has been shown that asynchronous computation helps some algorithms converge faster than the BSP system~\cite{graphlab}.

Another way to deal with the large amount of graph data is to use a graph-optimized database, such as neo4j~\cite{neo4j}. 
Neo4j provides an ACID-compliant transactional interface to a graph database. 
It also provides a declarative programming language called Cypher, so that algorithms expressed via Cypher can be executed natively inside the database.
Thus, neo4j makes it very easy for users to issue queries on large graphs.



\subsection{Flash Storage}
Many researchers have explored the use of secondary storage devices such as flash storage to counter the cost of large clusters with lots of DRAM, overcoming many of the challenges of using secondary storage for computation.
While the slow performance and long seek latency of disks prohibited serious use of external algorithms, flash storage is orders of magnitude faster both in terms of bandwidth and latency. 
However, flash storage still poses serious challenges in developing external algorithms, including still a much higher latency, in the range of 100s of microseconds compared to tens of nanoseconds of DRAM, and a coarser access granularity of 4KB to 8KB \emph{pages}, compared to DRAM's cache line.
Flash also suffers from lifetime issues due to erase and write causing physical wear in its storage cells.
Commercial flash storage devices handle wearing using an intermediate software called the Flash Translation Layer (FTL), which tries to assign writes to lesser-used cells, in a technique called \emph{wear leveling}.
Flash cell management is made even more difficult by the requirement that before pages can be written to, it must be \emph{erased}, an expensive operation of even coarser \emph{block} granularity of megabytes.
Random updates handled naively will cause a block erase every time, requiring all non-updated data to be read and re-written to the erased block.
Modern systems try to avoid this using many techniques including log-structured file systems~\cite{logfs}, but random updates are often still expensive operations.

\section{Sort-Reduce Method of Sequentializing Access}
\label{sec:sortreduce}

A generic vertex program in the push-style paradigm can be expressed as multiple iterations of the superstep shown in Algorithm~\ref{alg:baseline_vertex}.
The most intensive portion of this algorithm is the computation stage.
During the computation step, an \emph{edge program} is applied to each outgoing edge of each active vertex, and its result is used to update the value of the neighboring vertex, $V$.val, using a \emph{vertex program}.
Typically, both the current vertex value and the new value for the vertex computed during the superstep are maintained simultaneously and swapped at the end of the superstep. 
Both the old and the newly computed vertex values are needed to determine if the vertex remains active for the next superstep.
Note that the edge weights $E$.prop are constants in vertex programs.

\begin{algorithm}
\begin{algorithmic}
    \State $\triangleright$ Computation
	\For{each vertex $U$ in $active\_list$}
		\For{each edge $E(U,V)$}
			\State tmp = \textbf{edge\_program}($U$.val, $E$.prop)
			\State $V$.new\_val = \textbf{vertex\_program}($V$.new\_val, tmp)
		\EndFor
	\EndFor
    \State $\triangleright$ Update
    \State $active\_list$ = []
    \For{each vertex $V$}
        \If{\textbf{is\_active}($V$.val, $V$.new\_val)}
            \State $active\_list$.append($V$)
        \EndIf
        \State $V$.val = \textbf{finalize}($V$.val, $V$.new\_val)
        \State $\triangleright$ Initialization for next superstep
        \State $V$.new\_val = $V$.val
    \EndFor
\end{algorithmic}
\caption{Baseline vertex program superstep. Requires random read-modify-updates of $V$.val.}
\label{alg:baseline_vertex}
\end{algorithm}

This formulation can be used to express a wide range of algorithms by defining \texttt{edge\_program}, \texttt{vertex\_program}, \texttt{is\_active} and \texttt{finalize}.
For example, single-source-shortest-path can expressed by using addition as the edge program and the min function as the vertex program and the finalize function.
The active list is populated with vertices that saw their distance decrease in the last superstep (i.e. $V$.new\_val $<$ $V$.val) 

\begin{figure}[t]
	\begin{center}
	\includegraphics[page=1,width=0.48\textwidth]{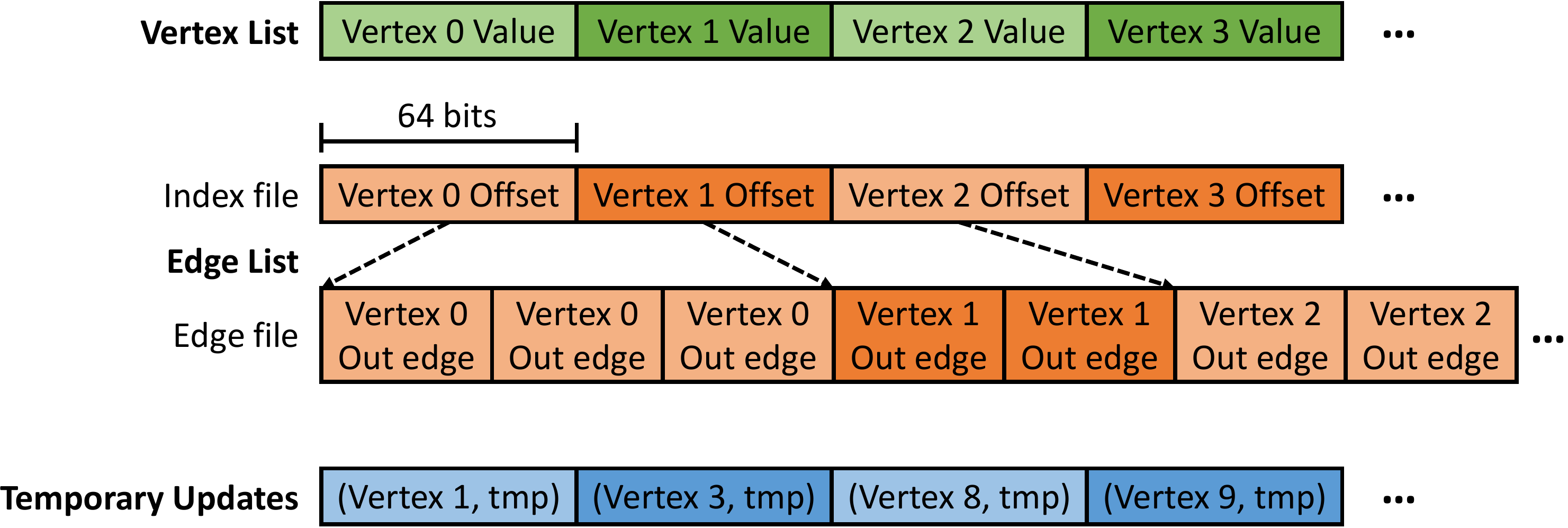}
	\caption{Storage used in BigSparse for Sort-Reduce}
    \label{fig:sort-reduce-storage}
	\end{center}
	\vspace{-10pt}
\end{figure}

Due to the irregular nature of sparse graphs, there is always a portion of the memory accesses in the computation stage of Algorithm~\ref{alg:baseline_vertex} that incurs fine-grained random accesses no matter which storage format is chosen for the edges and vertices.
This work assumes vertex data is stored sequentially and edge data is stored sorted by source vertex as seen in Figure~\ref{fig:sort-reduce-storage}.
With this storage format, a sorted active list will result in sequential accesses of edge data but fine-grained random accesses of destination vertex data.
These fine-grained random accesses are an especially big problem when accessing storage with a large access granularity such as Flash.

One way to eliminate random accesses from this formula is to collect all intermediate results from the \textbf{edge\_program} in the form of key-value pairs of $\langle V, intermediate\rangle$ as seen in Figure~\ref{fig:sort-reduce-storage}, and sort them by target destination $V$ before applying, as described in Algorithm~\ref{alg:sort_vertex}.
With a sorted list of updates, the target destinations are accessed in increasing order, and since sorting can be performed with sequential access, this method removes all fine-grained random access.
If the active list is sparse, there still may be some fine-grained accesses into the edge file, but since the accesses are still in increasing order, no pages in the edge file will incur the overhead of being read twice.
When the active list is dense, the size of the intermediate list becomes large and must be stored in secondary storage. 
As a result, the external sorting overhead can become large.

\begin{algorithm}
\begin{algorithmic}
    \State $\triangleright$ Computation
	\For{each vertex $U$ in $active\_list$}
		\For{each edge $E(U,V)$}
			\State tmp = \textbf{edge\_program}($U$.val, $E$.prop)
			\State $list$.add($\langle V, tmp\rangle$)
		\EndFor
	\EndFor
	\State $slist$ = sort($list$) by $V$
	\For{each pair $\langle V, tmp\rangle$ in $slist$}
		\State $V$.new\_val = \textbf{vertex\_program}($V$.new\_val, tmp)
	\EndFor
    \State \emph{Update omitted}
\end{algorithmic}
\caption{Computation portion of vertex program superstep that sorts vertex updates before applying the vertex program. This removes random accesses of $V$.val.}
\label{alg:sort_vertex}
\end{algorithm}

\begin{figure}[b]
	\centering
	\hspace*{\fill}%
	\subfloat[Completely sorting before reducing] {
		\includegraphics[page=1,width=0.2\textwidth]{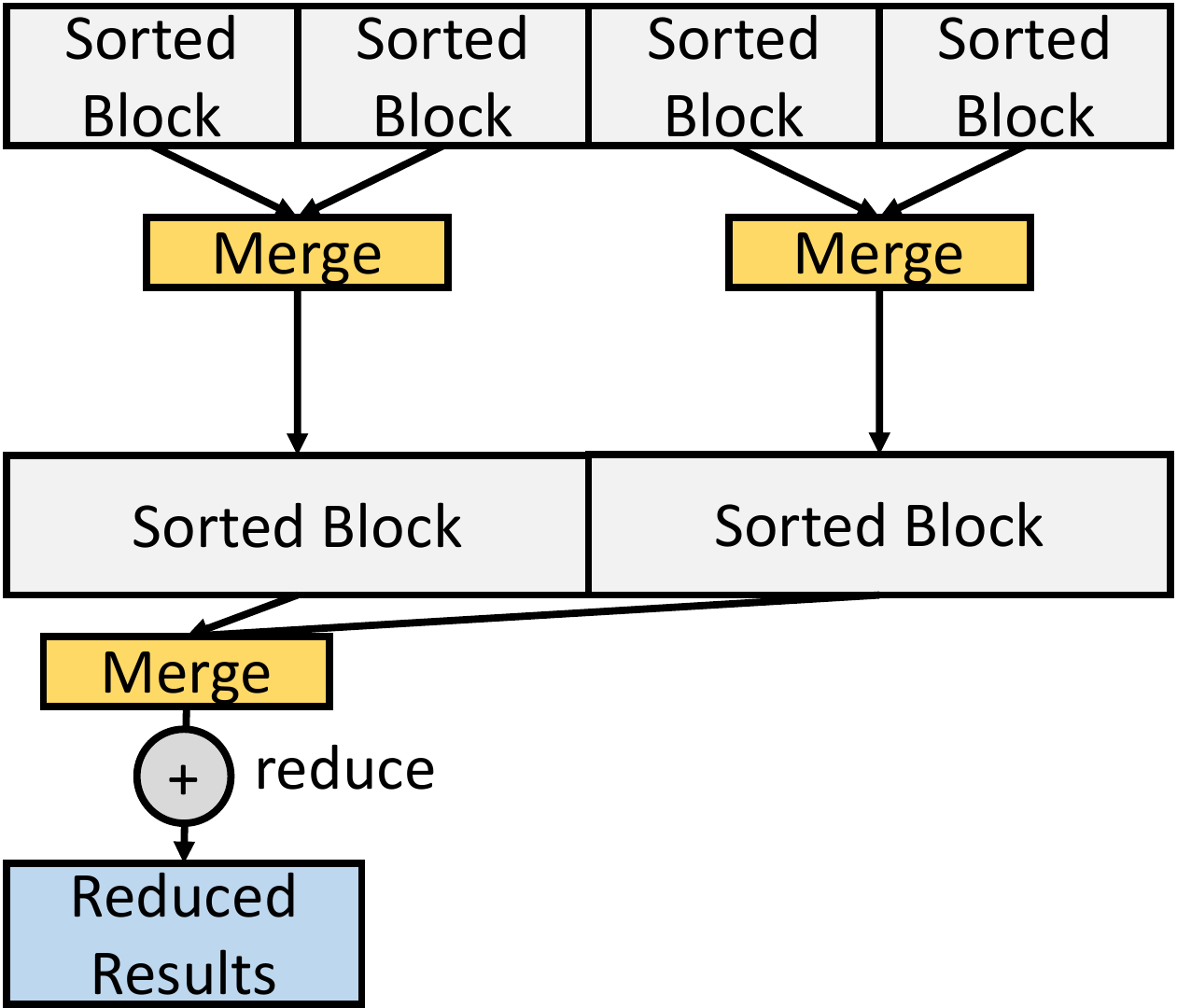}
		\label{fig:sort-conventional}
	}
	\hfill
	\subfloat[Interleaving sorting and reducing] {
		\includegraphics[page=2,width=0.2\textwidth]{figures/figures-crop.pdf}
		\label{fig:sort-interleaved}
	}
	\hspace*{\fill}%

	\caption{Interleaving sorting and reducing drastically reduces the amount of intermediate data to be sorted. This figure is drawn with $k$-to-1 mergers for $k$ = 2.}
	\label{fig:sort-and-reduce}
\end{figure}

Our BigSparse Architecture addresses the issue of fine-grained random accesses using a method we call Sort-Reduce.
The Sort-Reduce method logs and sorts all partial updates as in Algorithm~\ref{alg:sort_vertex}, but to cut down the overhead of sorting, updates to the same vertex are combined during the sorting process.
The temporary update list is first split into $n$-element blocks, and each of these blocks is sorted one-by-one.
Next there is a reduce step which combines updates to the same vertex in the same sorted block into a single update using the vertex program.
After that the sorting continues using a merge sort which merges $k$ blocks together and performs another reduce step.
This process iteratively reduces the amount of data to sort at each stage, which therefore reduces the number of secondary storage read/write accesses and improves performance.
The full Sort-Reduce method is shown in Algorithm~\ref{alg:sort_reduce}, and the reduction of data through two iterations of merging and reduction (with $k$ = 2) can be seen in Figure~\ref{fig:sort-and-reduce}.

\begin{algorithm}[t]
\begin{algorithmic}
    \State $\triangleright$ Phase 1 - Initial Update List Generation
	\For{each vertex $U$ in $active\_list$}
		\For{each edge $E(U,V)$}
            \State tmp = \textbf{edge\_program}($vertex\_values[U]$, $E$.prop)
			\State $list$.add($\langle V, tmp\rangle$)
		\EndFor
	\EndFor
    \For{each $n$ block $B$ in $list$}
        \State $sblock$ = sort($B$) by $V$
        \State $sblist$.append(reduce($sblock$) using
        \State the \textbf{vertex\_program})
    \EndFor
    \State $\triangleright$ Phase 2 - External Merge-Reduce
    \While{not\_sorted($sblist$)}
        \State $new\_sblist$ = []
        \For{each $k$ sorted blocks [$B_1$, $\ldots$, $B_k$] in $sblist$}
            \State sorted\_block = merge([$B_1$, $\ldots$, $B_k$])
            \State reduced\_sorted\_block = reduce(sorted\_block) using
            \State the \textbf{vertex\_program}
            \State $new\_sblist$.append(reduced\_sorted\_block)
        \EndFor
        \State $sblist$ = $new\_sblist$
    \EndWhile
    \State $\triangleright$ Determining the new active list and updates
    \State $active\_list$ = []
    \For{each pair $<V, tmp>$ in $sblist$}
        \State new\_val = \textbf{finalize}($V$.val, $tmp$)
        \If{\textbf{is\_active}($V$.val, new\_val)}
            \State $active\_list$.append($V$)
        \EndIf
        \State $V$.val = new\_val
    \EndFor
\end{algorithmic}
\caption{The Sort-Reduce Algorithm for executing a vertex program superstep. Interleaving reduction with sorting reduces the overhead of sorting.}
\label{alg:sort_reduce}
\end{algorithm}

In order to combine updates to the same vertex, BigSparse requires the vertex program to be a binary associative operation, i.e.:
$$vertex\_program(vertex\_program(A, B), C)$$ 
equals 
$$vertex\_program(A, vertex\_program(B, C))$$
This restriction is shared by other linear algebraic graph analytics systems, and is general enough to efficiently express many important graph algorithms of interest.

Another point to note about Algorithm~\ref{alg:sort_reduce} is that instead of updating $V$.val after each superstep, we can maintain the list of updates, i.e., $update\_list$.append($V$, $update\_val$), generated by each superstep. 
When in the next superstep the active vertex list is needed, the active vertices and the value of $U$.val can be computed on-the-fly by using each of the update lists in order.
Of course one can merge these update lists anytime if the access overhead becomes burdensome. 
Such a scheme can be a win for flash storage where there is no in situ update of data and the inaccessible data is garbage collected in background.

There are many choices to make when implementing the Sort-Reduce method, including parameters $n$ and $k$ and how tasks are parallelized across threads.
The next section introduces the BigSparse Architecture and explains how it selects $n$ and $k$ and parallelizes the Sort-Reduce method in order to maintain full utilization of the Flash bandwidth.

\section{BigSparse Architecture}
\label{sec:design}

The BigSparse architecture leverages the Sort-Reduce method to produce efficient, parallelized, external implementations of user-defined graph algorithms.
The vertex and edge data for the graph are stored in a compact compressed format in secondary storage, preferably using an array of high-performance flash storage (SSDs).
As discussed earlier, the use of secondary storage presents two major problems in the implementation of graph algorithms: large access granularity and long access latency.
The Sort-Reduce Algorithm~\ref{alg:sort_reduce} mitigates some of the effects of large access granularity, but it is the responsibility of the BigSparse architecture's implementation to be able to tolerate the long access latency and to keep the secondary storage's bandwidth saturated while minimizing DRAM usage.

We first describe the overall stream-processing architecture for processing one superstep of the algorithm.
Next we introduce the storage system and the data representation.
Then we give the details of each of the computational blocks of this architecture.
Finally we discuss the choice of parameters in the architecture, and how they should change according the storage's bandwidth and latency.

\subsection{Architecture Overview of a Superstep}

\begin{figure}[h]
	\begin{center}
	\subfloat[Phase 1]{\includegraphics[page=4,width=0.4\textwidth]{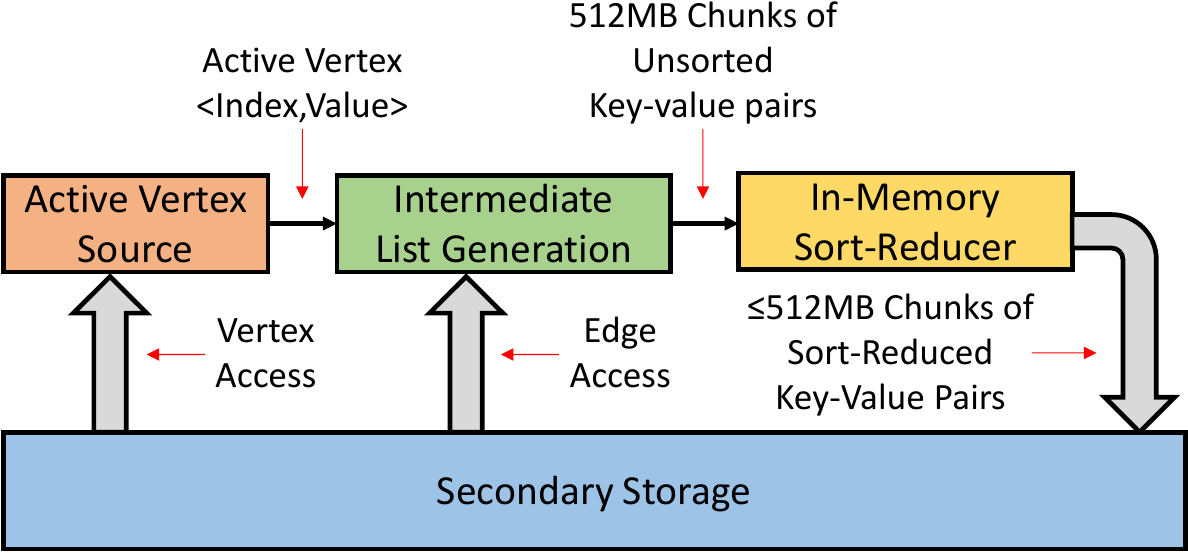}} \\
	\subfloat[Phase 2]{\includegraphics[page=6,width=0.4\textwidth]{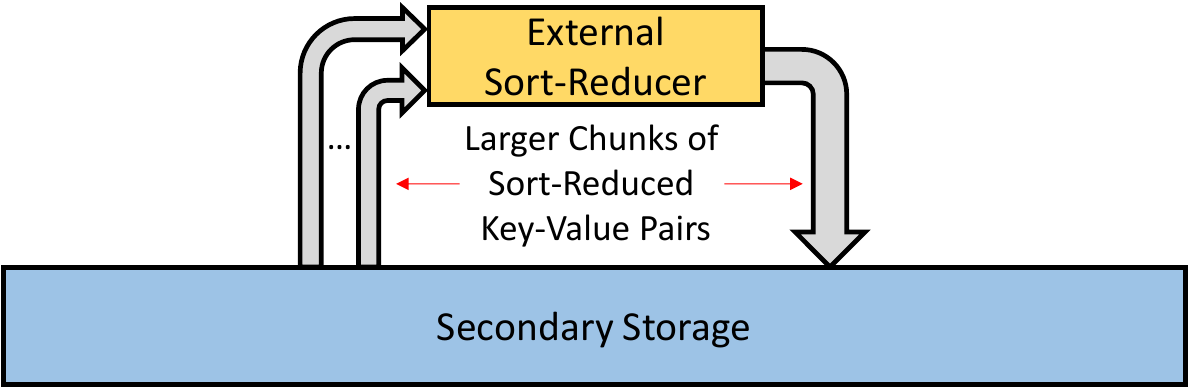}}
	\caption{The two phases of a graph algorithm superstep in BigSparse.}
	\label{fig:graphproc-overview}
	\end{center}
	\vspace{-10pt}
\end{figure}

The execution of a superstep on the BigSparse architecture can be viewed in two phases where phase two is executed iteratively until all vertices have been updated in the storage (see Figure~\ref{fig:graphproc-overview}).
Each phase constitutes a software pipeline with explicit sequential accesses to secondary storage.
Each computational block in the figure is implemented with multiple threads.
Thick arrows in the figure are sequential secondary storage accesses, while thin arrows are in-memory FIFO buffers.

The secondary storage block is an abstraction for the array of secondary storage devices (SSDs) where each storage device is running its own file system.
BigSparse software stripes its blocks across files, and uses Linux's asynchronous IO interface for parallel IO accesses by the same thread.

Given an active vertex list, phase 1 reads enough vertex and edge data from the storage to produce 512 MB chunks and feeds them into the in-memory sort reducer which in turn produces a sorted (and reduced) chunk of less than or equal to 512 MB. 
The intermediate list generation block in the figure embodies the edge program, while the in-memory sort-reducer block uses the vertex program for reduction.
Phase 1 computation, because of its streaming nature, only requires enough DRAM to process these 512 MB chunks. 
In order to saturate the storage bandwidth, it also requires DRAM IO buffers to hide secondary storage latency.

Phase 2 performs external $k$-to-1 merge-sorting and reduction (using the vertex program) on k blocks in a streaming manner, and stores the larger sorted blocks back to storage for the next iteration of phase 2.
This phase streams from k distinct blocks at a time, but all these streams share the same secondary storage hardware in a highly interleaved manner.
The number of phase 2 iterations required is proportional to $log_k(|initial\_update\_list|)$, and in the current implementation there is no overlap in the execution of iterations or phases.
After the final external sort-reduce iteration, a new active list and updated vertex values are prepared for the next superstep by configuring the active vertex source block of the next superstep.
The configuration of this block requires no accesses to secondary storage; instead it is configured to compute the new active list and vertex values just-in-time provided logs of updates for past supersteps.

\subsection{Storage and Data Representation}
\subsubsection{Storage Management using Append-Only Files}

\begin{figure}[tb]
	\begin{center}
	\includegraphics[page=3,width=0.3\textwidth]{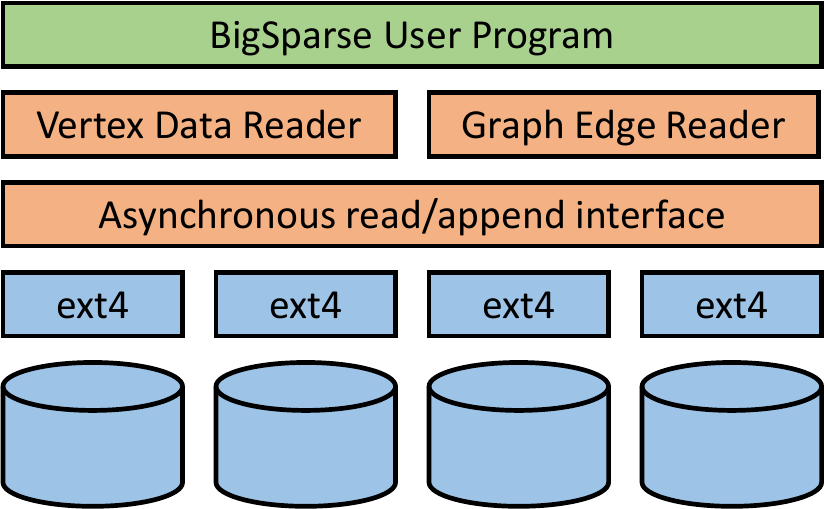}
	\caption{Storage access hierarchy in a BigSparse system}
	\label{fig:storage-management}
	\end{center}
	\vspace{-10pt}
\end{figure}

BigSparse operates on an array of secondary storage devices plugged into a multicore host server.
Each storage devices is formatted with its own file system such as ext4, which make it easy for BigSparse to share devices with other programs and the operating system.
The overall storage management hierarchy of BigSparse can be seen in Figure~\ref{fig:storage-management}.

BigSparse manages all files in an append-only manner, meaning data can only be appended to the end of a file and random updates are disallowed.
This append-only behavior is sufficient for the BigSparse architecture since all writes to storage are sequential.
Using append-only mode for BigSparse files also simplifies the work of the Flash Translation Layer (FTL) within SSDs, specifically wear leveling and page allocation.
This also helps extend flash lifetime.

The storage management layer exposes an asynchronous read and append interfaces to the rest of the software platform, which allows upper layers to extract high performance from storage without spawning multiple threads just to issue read/write requests.
Beneath this layer, BigSparse manages its own data distribution using a RAID0-like striping scheme over its array of storage devices.
Data is split into chunks of 1MB, and stored in each storage device in a round-robin fashion.
At the start of program execution, BigSparse spawns read and append threads for each storage device in its array.
These threads maintain in-order operational semantics for read and append operations, therefore simplifying the upper level software development.

\subsubsection{Graph Data Representation}

Due to the lower bandwidth of secondary storage compared to DRAM, it is important to use storage efficiently to increase its effective bandwidth (e.g. edges read per second).
Therefore, BigSparse stores graphs in a compact binary encoded format: all edges are laid out in order of source vertex in a single \emph{edge file} in storage, and each edge value contains the destination vertex index and the edge value.
An \emph{index file} contains a 64-bit byte offset per each source vertex which points to the region of the edge file containing outgoing edges for that vertex.
Using a byte offset instead of item index allows variable length edge attributes, including strings and arrays.
The relationship between the index file and the edge file can be seen in Figure~\ref{fig:sort-reduce-storage}.




\subsubsection{Vertex Data Representation}

Vertex data is stored in secondary storage as either densely or sparsely encoded files. Densely encoded files have fixed sizes occording to the number of vertices in a graph, and can be randomly addressed using the vertex index as key. Sparsely encoded files are sorted lists of pairs of vertex indices and values.
In all of BigSparse operation, sparse vertex lists are accessed and used in a sequential fashion, starting from the beginning and proceeding until data from the particular vertex file is no longer needed.
This characteristic allows deep prefetching of vertex files.

Unlike graph data, there is not a one-to-one correspondence between vertex data stored in secondary storage and vertex data used by the BigSparse architecture.
Between supersteps of a graph algorithm, the vertex values for a graph may be spread across many vertex files: one file containing initial values and the other files containing incremental updates.
Likewise an active vertex list may be stored as references to multiple vertex files along with operations to determine which vertices in the graph are active and calculate the vertex value of the active vertices.
During the operation of BigSparse, vertex files can be merged and consolidated to simplify future accesses of vertex values or active lists.

\subsection{Active Vertex Source}
The active vertex source block reads from one or more vertex files in storage and streams the desired active vertex data to the next block for intermediate list generation.
The active vertex list is computed just-in-time to avoid the overhead of additional storage access.
For breadth-first-search, the just-in-time computation is used to get the vertex values involves streaming through vertex files of updates from each superstep to find the most recent update for each vertex.
For PageRank, since all the vertices are active each superstep, the active vertex source just applies a dampening factor to the most recent set of updates to get the new active list and vertex values.

The active vertex source block is constructed by connecting multiple \emph{vertex source} objects together in a dataflow fashion.
Each vertex source interface exposes three functions: \texttt{HasNext(\textit{idx=none})}, \texttt{GetNext()} and \texttt{Rewind()}.
A vertex source streams through all vertex values in itself from begining to end.
BigSparse provides different types of vertex sources that implement this interface.
When \texttt{HasNext} is provided with a vertex index, it either fast-forwards until a vertex of that index is found, or if the vertex source is a densely encoded file, jumps directly to that location in the file.
\emph{Compound vertex sources} take one or more vertex sources as input and return transformed vertex data using either arithmetic, logical, or set operations.
All compound vertex sources are evaluated just-in-time at runtime, when \texttt{HasNext} and \texttt{GetNext} are called, to avoid updates on storage.
A compound vertex source can use \texttt{HasNext} with an index argument to efficiently process sparse vertex lists.
Here is the list of available vertex sources:

\begin{enumerate}
\setlength\itemsep{.1em}
\item \texttt{VListFileSparse}: Reads vertex data from a sparse vertex file in storage.
\vspace{-2pt}
\item \texttt{VListFileDense}: Reads vertex data from a dense vertex file in storage. A bloom filter reduces unnecessary accesses.
\vspace{-2pt}
\item \texttt{VListArith}: Takes a vertex source $S$ and a function pointer $arithFunc$ as input, and for each $\langle K, V\rangle$ in $S$, returns $\langle K, arithFunct(V)\rangle$.
\vspace{-2pt}
\item \texttt{VListLogical}: Takes two vertex sources as input and performs set operations on them, some of which are provided in Table~\ref{tab:vlist_logical}.
\vspace{-2pt}
\item \texttt{VListSplit}: Takes a vertex source $S$ and distributes it across multiple vertex source interfaces. Vertices are distributed in chunks of $N$ to each of the output interfaces. A large $N$ is used for cache efficiency. This is used to partition the graph for parallelization.
\vspace{-2pt}
\end{enumerate}

\begin{table}[b]
\begin{tabularx}{0.5\textwidth}{l|X}
\hline
Name & Description \\
\hline
\hline
Union & Set union of sources. Takes value from first source if keys match \\
Difference & Set difference of sources \\
Minimum & Element-wise minimum between sources, returns only when keys match\\
Custom & Uses custom compute function when keys match to determine value \\
Converge & Takes custom compare function and returns vertices that have not converged\\
\hline
\end{tabularx}
\caption{Logical operations supported by VListLogical}
\label{tab:vlist_logical}
\vspace{-10pt}
\end{table}

\subsection{Intermediate List Generation}

\begin{figure}[h]
	\begin{center}
	\includegraphics[page=5,width=0.5\textwidth]{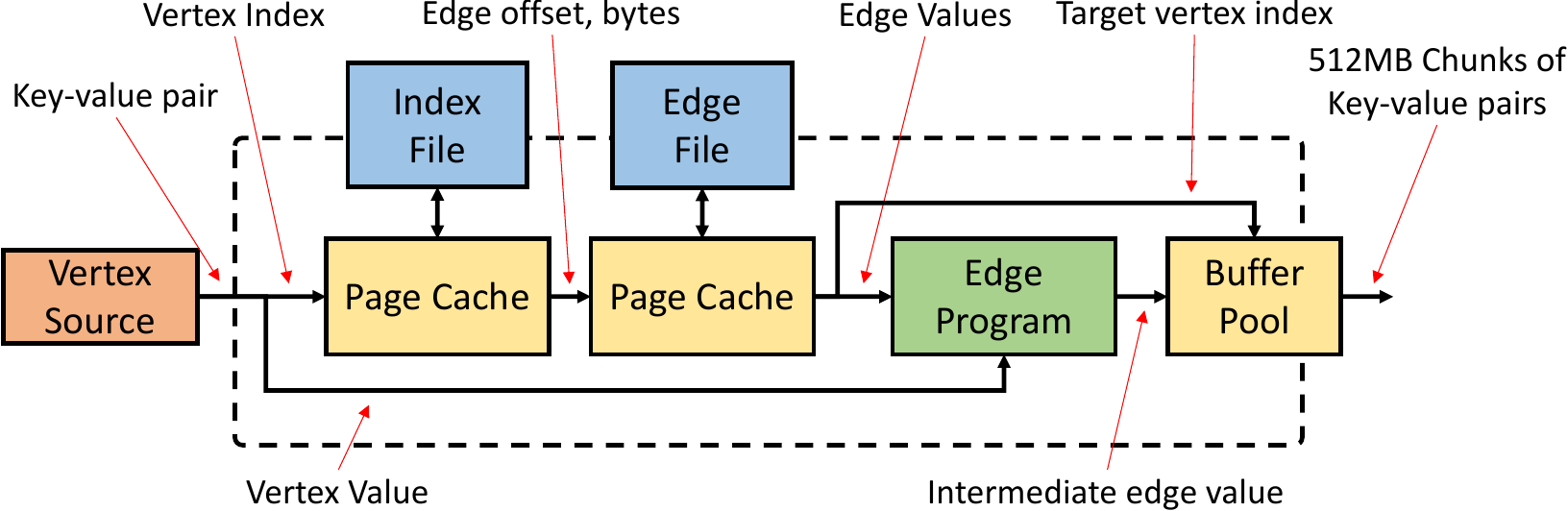}
	\caption{Intermediate list generation dataflow in the edge processor module}
	\label{fig:edgeproc-internal}
	\end{center}
	\vspace{-10pt}
\end{figure}

The intermediate list generation component reads active vertex values from the active vertex source, applies the edge program to its outgoing edges, and streams the partial update results in 512 MB chunks to the in-memory sort-reduce block.
Figure~\ref{fig:edgeproc-internal} describes the dataflow of intermediate list generation.
All storage accesses are asynchronous, and the dataflow is pipelined, so that there are thousands of vertices being processed at any time.

Outgoing edge data is read for each active vertex by first accessing the index file, and then accessing the edge file.
Since the vertex indices from the active vertex source are increasing, the addresses accessed in the two files are also always increasing.
As a result, all accesses to a particular page happen back-to-back.
BigSparse maintains a page cache containing the last page accessed from the index and edge files so the same page never has to be fetched from storage twice during the same superstep.

The results of the edge program is coupled with the destination vertex index from the edge file, and logged.
The edge processor module maintains a pool of 512MB buffers, and a buffer is emitted into a queue whenever it is filled.
These 512MB buffers are used during the sort-reduce process, and are returned to the buffer pool after they are used.
The size of the 512MB buffer was chosen because of the balance between computation cost and memory capacity, and will be explained in more detail in the following section.

\begin{figure}[h]
	\begin{center}
	\includegraphics[page=7,width=0.4\textwidth]{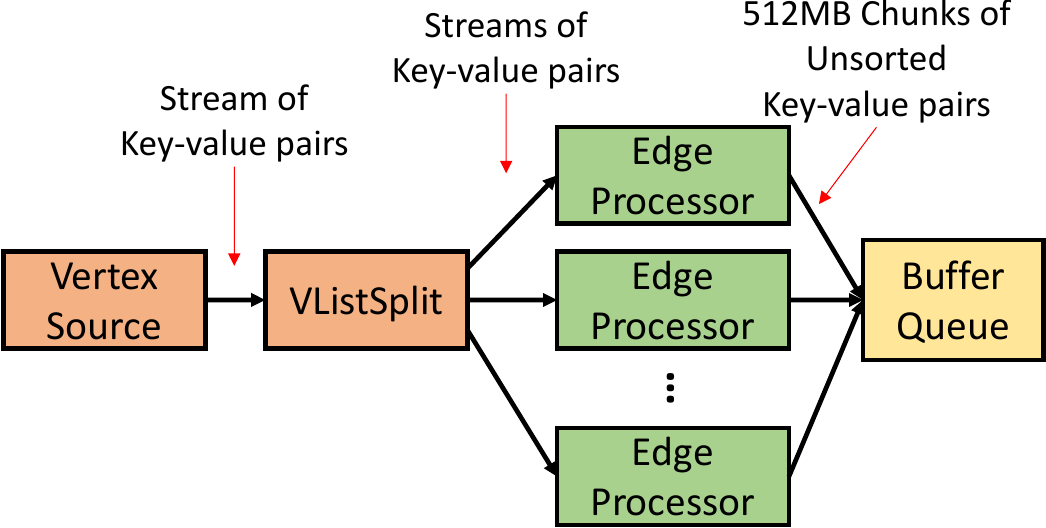}
	\caption{Multiple edge process threads work in parallel using the VListSplit vertex source}
	\label{fig:edgeproc-process}
	\end{center}
	\vspace{-10pt}
\end{figure}

As secondary storage devices become faster, the edge processor module's data marshalling overhead may become the bottleneck instead of storage access.
BigSparse solves this problem by using the VListSplit vertex data source to partition the input vertex data set, and parallelize work across multiple threads of edge processors as seen in Figure~\ref{fig:edgeproc-process}.
The VListSplit source is configured to read millions of vertices at a time before handing the token over to one of the threads of intermediate list generation.
The results from each edge processor is collected into a queue of buffers for the in-memory sort-reduce block to use.

\subsection{In-Memory Sort-Reduce}

\begin{figure}[hb]
	\begin{center}
	\includegraphics[page=7,width=0.4\textwidth]{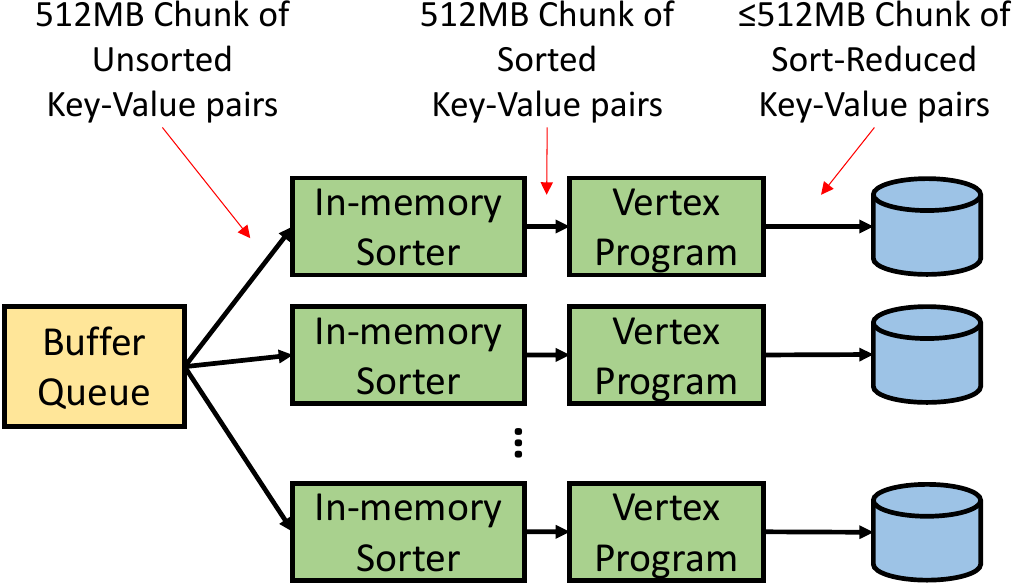}
	\caption{The first phase of sort-reduce happens in in-memory chunks}
	\label{fig:inmem-sorter}
	\end{center}
\end{figure}

The in-memory sort-reduce block takes 512~MB blocks of partial updates from the intermediate list generation block, and then sorts and reduces them to produce a chunk of sorted intermediate updates less than or equal 512~MB to write to storage in preparation for the external sort-reduce block in phase 2.
The overall dataflow of the in-memory sort-reducer can be seen in Figure~\ref{fig:inmem-sorter}. 
This block is in a pipelined fashion along with the other blocks of phase 1.
The output of this block is a stream of temporary files that are lists of partial updates sorted by vertex index.

The sorter used in the in-memory sorter is a multithread implementation, which first sorts 32~MB sub-chunks using quicksort, and then performs a merge sort using a multithreaded merger.
The same merger is used for the external sort-reducer in phase 2.

Once the resulting chunk is stored in a file in secondary storage, The file handle for the new file is pushed into a work queue for the phase 2 external sort-reducer to manage, and the used buffer is returned to the edge processor's buffer pool in Figure~\ref{fig:edgeproc-internal}.
Because sorting the entirety of the 512MB chunk is time-consuming, multiple in-memory sort-reduce threads are spawned, and each thread creates their own output files.

\subsection{External Sort-Reduce}

\begin{figure}[hb]
	\begin{center}
	\includegraphics[page=8,width=0.5\textwidth]{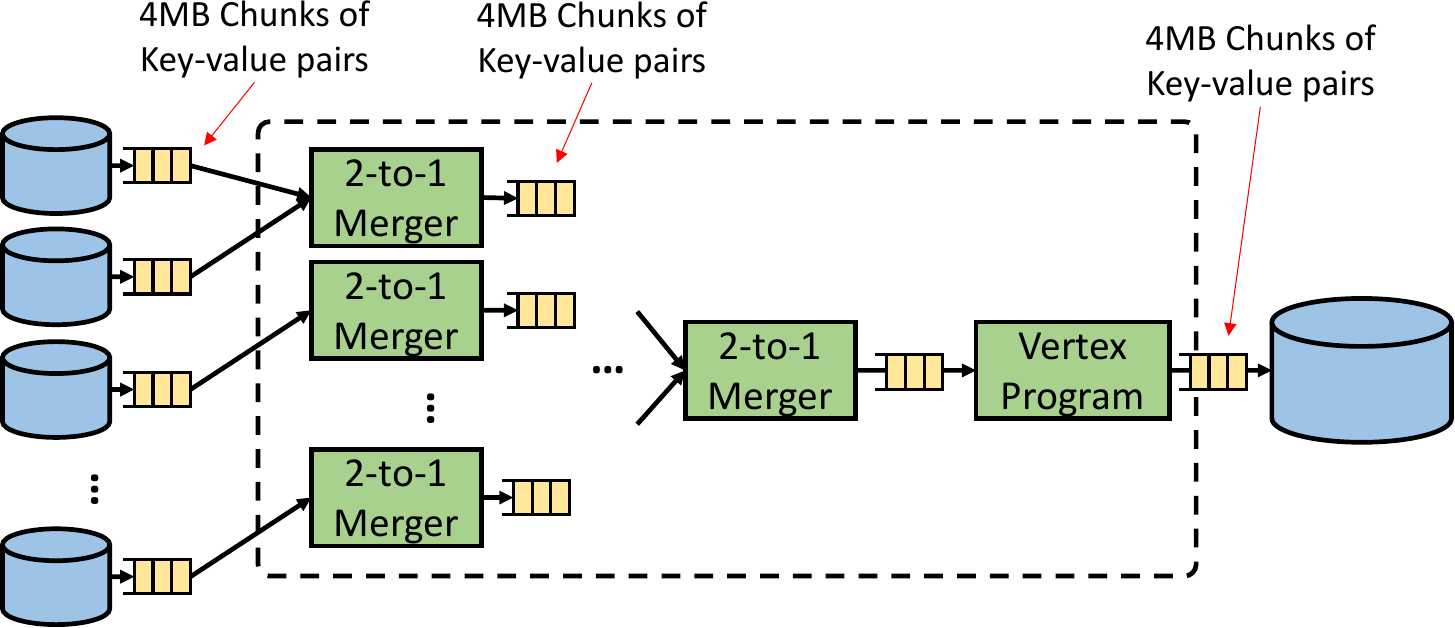}
	\caption{Merge-reducer progressively performs $k$-to-1 merge and reduce before writing to disk}
	\label{fig:merge-reducer}
	\end{center}
	\vspace{-10pt}
\end{figure}

External sort-reduce performs sort-reduce on partially sort-reduced files generated by either the in-memory or external sort-reducers, until all data is reduced into a single file, which is the final output of the algorithm.
External sort-reduce is performed by a pool of $k$-to-1 merge-reducer modules.
The structure of the $k$-to-1 merge-reducer module can be seen in Figure~\ref{fig:merge-reducer}.

Each chunk of sort-reduced data is stored in its own file.
The External Sort-Reduce takes $k$ files and merges them together and then performs reduction on matching vertex indices using the vertex program.
The current list of temporary files are kept in a work queue, and a free merge-reducer is invoked whenever there are more than one files in the queue and a free merger exists.
After merge-reduced data is written to a new file, its file handle is pushed into the work queue.
This process is repeated until there is only one file in the work queue.

The size of the pool of merge-reduce networks depends on the number of available CPU cores on the server and the bandwidth of secondary storage, as it would be meaningless to have more mergers if there is either not enough processing power, or not enough storage bandwidth.

The merger prefetches 4MB file chunks using the asynchronous file interface to keep the storage and the input buffer queue busy.
The $k$-to-1 merge-reducer is internally implemented with many 2-to-1 mergers organized into a tree-like hierarchy.
To make the best use of multicore CPUs, each 2-to-1 merger runs on its own thread and passes partially merged logs among each other in 4MB chunks.
2-1 mergers are used due to the limited single-core performance of CPUs; a single thread implementation of a $k$-to-1 merger quickly becomes the bottleneck even for a moderately fast SSD.
At the end of the merge hierarchy, the stream is reduced using the vertex program whenever adjacent log entries have matching keys.

\subsection{Architectural Tuning for Storage Efficiency}

The BigSparse system is highly parameterized for adjusting to different system configurations.
In most cases the secondary storage will be the limiting factor for performance, so the system parameters should be tuned to achieve maximum secondary storage bandwidth utilization.
Without efficient tuning, portions of the pipeline may get backed up or starved.
Backed up pipelines will reduce the rate at which storage is read, and starved pipelines will reduce the rate at which storage is written.
The values mentioned in this section, such as buffer sizes and merge-sort diameter, were chosen for various reasons, but all the choices come together to ensure that storage bandwidth is utilized as much as possible.

The in-memory sort buffer size is chosen as a result of memory and CPU performance limitations.
It is generally beneficial to have larger in-memory sorting units because we can do more computation in DRAM, instead of the slower secondary storage.
On the other hand, we must be able to fit enough buffers in memory to keep enough sorter threads busy.
The in-memory sorting overhead must also be balanced with the intermediate list generation rate, which ultimately depends on storage performance.
The buffer size can be dynamically made smaller or larger depending on the available allocated memory and processing power.
For our evaluation environment, the buffer size of 512MB was chosen as it will keep the memory requirement below 8GBs while also keeping the computation load for sorting below 16 threads, which we thought were reasonable resources for an affordable server node.

\section{Evaluation}
\label{sec:evaluation}

We evaluate the performance of BigSparse, compared to other in-memory, semi-external and external graph analytics systems to show that BigSparse becomes the desirable system when graph sizes become large relative to available hardware.
In order to demonstrate this, we evaluate the performance of various graph analytics software on graphs of different sizes, using different system configurations such as different storage capacity, bandwidth and memory capacity.
Our experiments can be categorized into two types: the experiments dealing with small graphs that can fit in memory of reasonable machines or clusters, and the experiments with large graphs that push the limit of software and hardware capacities.
We see that for small graphs, in-memory systems like GraphLab and Galois are the fastest, but they become inapplicable as problem size increases.
Semi-external systems such as FlashGraph provide excellent performance as long as the vertex data fits in memory.
However, as problem sizes increases even semi-external system performances drop catastrophically.
For such graphs, BigSparse is the only system that continues to deliver high performance.

\subsection{Experimental System Setup}

We used three different machine setups for our experiments.
We conducted small graph experiments on a single server with 32 cores of Intel Xeon E5-2690 (2.90GHz) and 128GBs of DRAM. It was equipped with a single 512GB PCIe SSD, providing up to 1.2GB/s sequential read bandwidth.
Large graph experiments were conducted with the same server, except that it was connected to five 512GB PCIe SSDs, adding up to 2.5TB of flash storage and approximately 6GB/s of sequential read bandwidth.
For the distributed GraphLab experiments, we used a small cluster of machines, each with 24 Xeon X5670 cores running at 2.93MHz, and equipped with 48GB of memory and networked via 1 gigabit ethernet.

We compared the performance of various in-memory, semi-external and fully external graph analytics software.
We ran algorithm implementations that were developed and distributed by their developers, and tried to find the best software configurations to get maximum performance out of them, experimenting with features such as buffer and cache size, and thread count.

\subsection{Graph Algorithms and Their Implementation}
\emph{\textbf{Breadth-First-Search: }}
Breadth-First-Search traverses a graph by iteratively visiting the immediate neighbors of the active vertices of the current iteration.
BFS maintains a parent node for each visited vertex, so that each vertex can be traced back to the root vertex.
This algorithm can be expressed using the following vertex program. $vertexID$ is provided by the system:

\begin{algorithmic}
\Function{edgeprog}{$vertexValue$,$edgeValue$,$vertexID$}
	\State \Return $vertexID$
\EndFunction
\Function{vertexprog}{$vertexValue1$,$vertexValue2$}
	\State \Return $vertexValue1$
\EndFunction
\end{algorithmic}

The edge program activates all neighboring nodes and sends them the vertex ID of itself as the parent.
The vertex program is called whenever a node has more than one parent, and selects the first parent that visited it.

The algorithm keeps track of a \emph{visited} vertex list using a densely encoded vertex source coupled with an in-memory filter.
After every iteration of the algorithm, the resulting list is checked against the visited list using the logical set operator. 
If the set difference between the result and \emph{visited} is an empty set, the algorithm terminates.
Otherwise, the set difference is used as the next active vertex list.
The visited list is updated using the set union operator.

BFS is a good example of an algorithm with sparse active vertices.
BFS is a very important algorithm because it forms the basis and shares the characteristics of many other algorithms such as Single-Source Shortest Path (SSSP) and Label Propagation.

\emph{\textbf{PageRank: }}
PageRank is Google's algorithm of determining the importance of web pages using its relationship with other pages.
PageRank starts with a vertex list with all vertex values initialized to $1/NumVertices$.
At every iteration, each vertex sends its neighbors its own value divided by the number of its neighbors.
The recipients add these values together and add a dampening factor using the following equation:
$0.15/NumVertices + 0.85*sum$.
The algorithm can be expressed using the following program. $numNeighbors$ is provided by the system:

\begin{algorithmic}
\Function{edgeprog}{$vertexValue$,$edgeValue$,$numNeighbor$}
	\State \Return $vertexValue / numNeighbor$
\EndFunction
\Function{vertexprog}{$vertexValue1$,$vertexValue2$}
	\State \Return $vertexValue1$+$vertexValue2$
\EndFunction
\end{algorithmic}

After each iteration of the algorithm, the dampening factor is applied to the resulting vertex list using a lazily evaluated preprocessor.
The algorithm keeps track of the current pagerank list \emph{currentRank} using a densely encoded vertex source and an in-memory filter.
After every iteration, the resulting vertex list is checked against the \emph{currentRank} list using the \emph{check convergence} set operation.
If the resulting vertex list is empty, the algorithm terminates.
Otherwise, the result is used as the active list for the next iteration, and the \emph{currentRank} list is updated using the set union operator.

PageRank is a good example of an algorithm with very dense active vertex sets.
In order to remove the performance effects of various algorithmic modifications to PageRank and do a fair comparison between systems, we only measured the performance of the very first iteration of PageRank, when all vertices are active.

\emph{\textbf{Betweenness-Centrality: }}
Betweenness-centrality calculates each node's centrality in the graph.
It is calculated by performing BFS from a vertex and then backtracing, in order to calculate the number of shortest paths that go through each vertex.
The edge and vertex programs for betweenness centrality is identical to BFS.

After traversal is finished, each generated vertex list's vertex values are each vertex's parent vertex ID.
Each list can be made ready for backtracing by taking the vertex values as keys, and sort-reducing them.
The backtrace sort-reduce algorithm can be expressed with the following algorithm:

\begin{algorithmic}
\Function{edgeprog}{$vertexValue$,$edgeValue$,$null$}
	\State \Return 1
\EndFunction
\Function{vertexprog}{$vertexValue1$,$vertexValue2$}
	\State \Return $vertexValue1$+$vertexValue2$
\EndFunction
\end{algorithmic}

Once all lists have been reversed and reduced, the final result can be calculated by applying set union to all reversed vertex lists using a cascade of set union operations, with a custom function to add multiply values whenever there is a key match.

BC is a good example of an algorithm that requires backtracing, which is an important tool for many applications including machine learning and bioinformatics.

%
%
%

\subsection{Graph Datasets}
For each application, we used multiple different graph datasets with different sizes. 
One important graph dataset we analyzed is the Web Data Commons (WDC) web crawl graph~\cite{wdcgraph} with over 3.5 billion vertices, adding up to over 2TBs in text format.
The WDC graph is one of the largest real-world graphs that is available.
Others include kronecker graphs generated at various sizes according to Graph 500 configurations, and the popular twitter graph.
Table~\ref{tab:dataset} describes some of the graphs of interest.
The size of the dataset is measured after column compressed binary encoding in BigSparse's format.

\begin{table}[h]
\centering
\begin{tabular}{|c||r|r|r|r|r}
\hline
name & twitter & kron28 & kron30 & wdc \\
\hline
nodes & 41M & 268M & 1B & 3B \\
edges & 1.47B &4B & 17B & 128B \\
size &  6GB & 18GB & 72GB & 502GB \\
txtsize & 25GB & 88GB & 351GB & 2648GB \\
\hline
\end{tabular}
\caption{Graph datasets that were examined}
\label{tab:dataset}
\end{table}

%
%
%


\subsection {Evaluation with Small Graphs (Less Than 1 Billion edges)}

\begin{figure}[t!]
	\centering
	\subfloat[Execution time of PageRank on small graphs using various systems] {
		\includegraphics[width=0.4\textwidth]{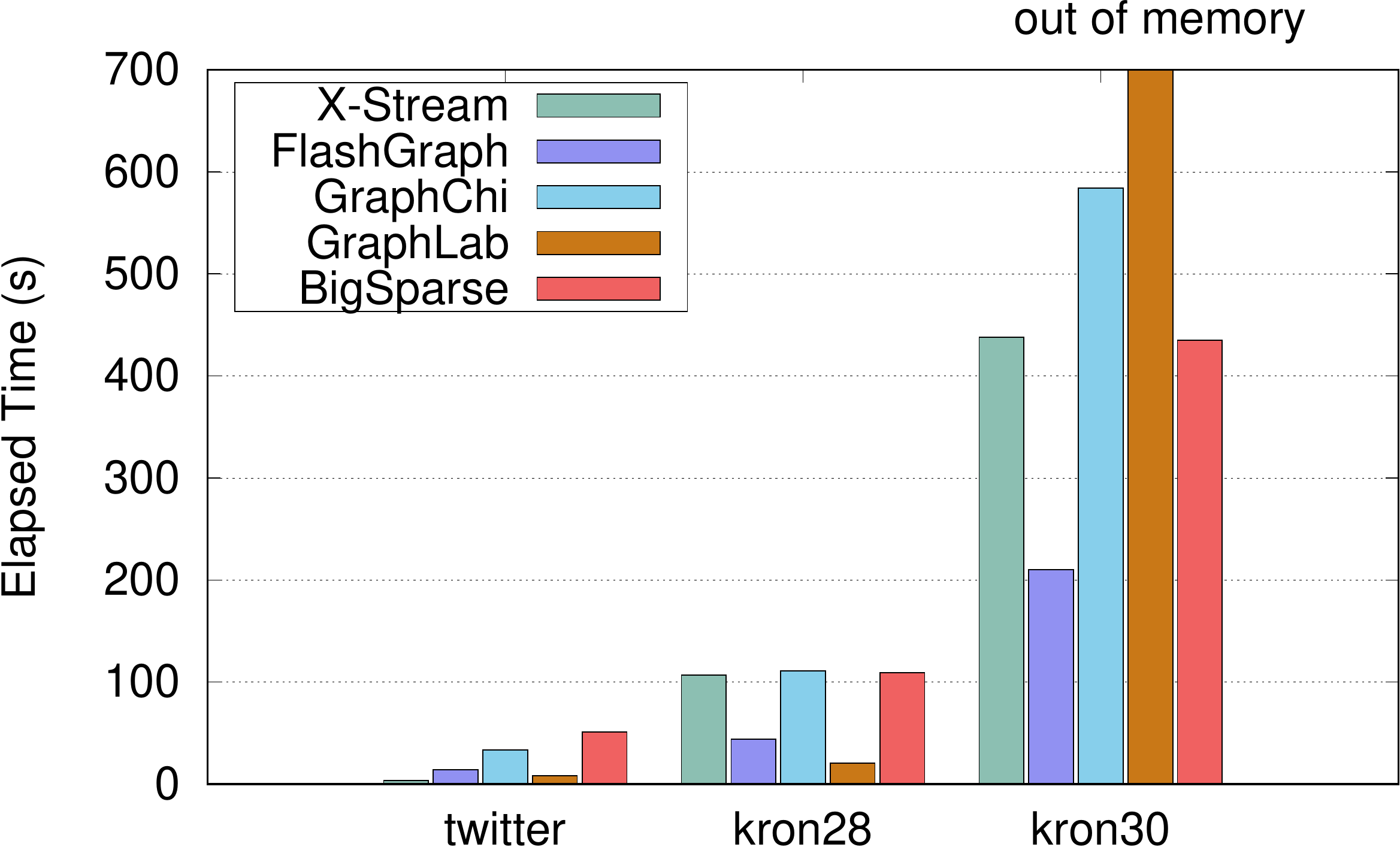}
		\label{fig:perf-pagerank-small}
		\vspace{-10pt}
	}
	\hfill
	\subfloat[Execution time of breadth first search on small graphs using various systems] {
		\includegraphics[width=0.4\textwidth]{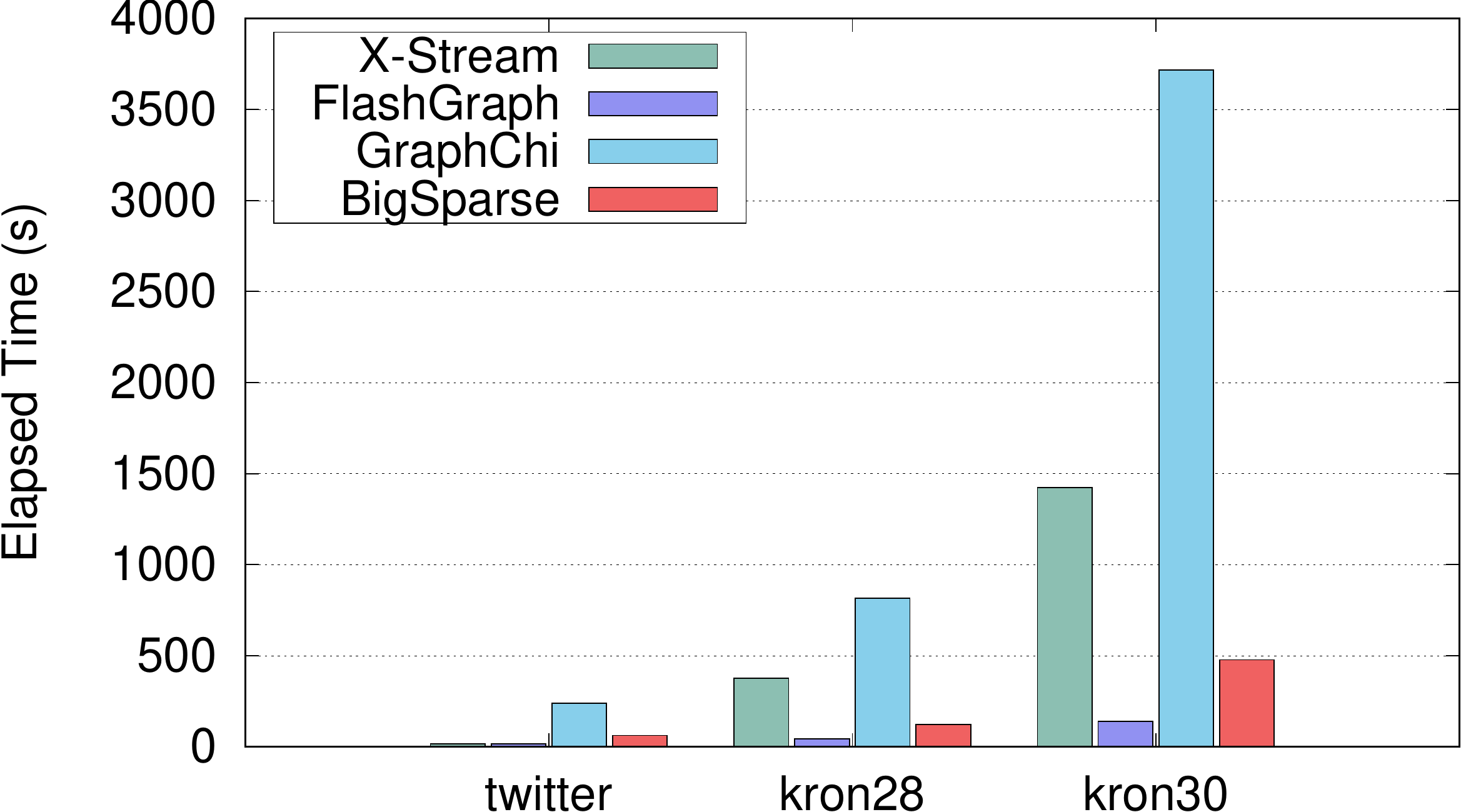}
		\label{fig:perf-bfs-small}
		\vspace{-10pt}
	}
	\hfill
	\subfloat[Execution time of betweenness centrality on small graphs using various systems] {
		\includegraphics[width=0.4\textwidth]{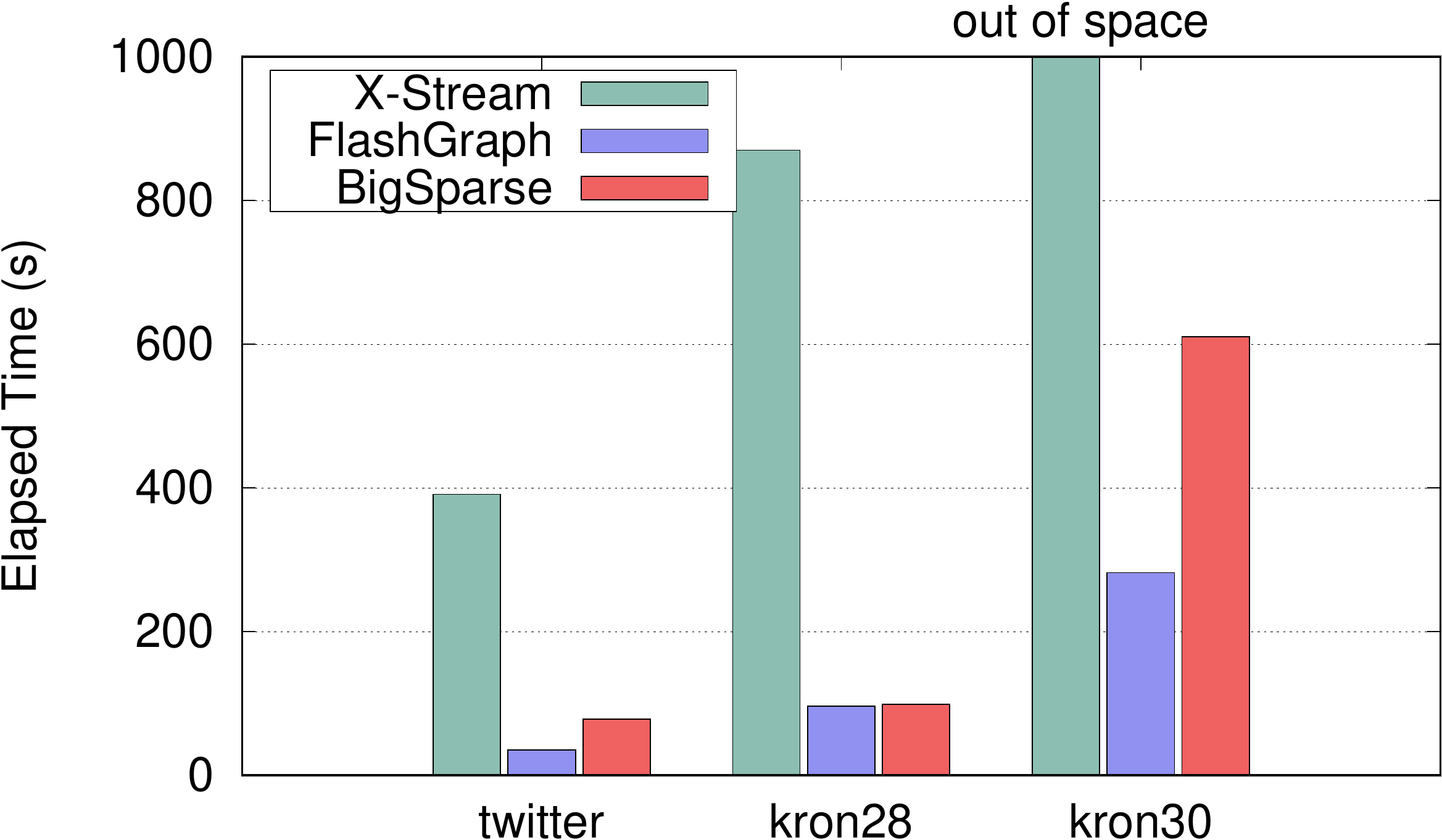}
		\label{fig:perf-bc-small}
		\vspace{-10pt}
	}
	\hfill 

	\caption{Execution time of graph algorithms on small graphs (On a single server with one SSD)}
	\label{fig:sort-compare}
	\vspace{-10pt}
\end{figure}

\subsubsection {PageRank Results} 
Figure~\ref{fig:perf-pagerank-small} shows the performance comparison of various systems running PageRank on small graphs.
All systems except GraphLab were executed on a single-node experimental machine setup described above.
GraphLab was distributed across a small cluster that was also described above.
GraphLab was run on a cluster of four machines for the twitter graph adding up to 96 cores and 192GB of memory, and on a cluster of five machines for the Kron28 graph, adding up to 120 cores and 240GB of memory.
We were unable to fit the Kron30 dataset on the cluster that we have, and is marked as \emph{out of memory}.

While the in-memory GraphLab is generally the fastest, it quickly becomes inapplicable because the collective memory capacity limits the size of graphs that can be processed.
As graphs become larger and in-memory systems fail, semi-external systems like FlashGraph becomes the most viable alternative.

%
%

\subsubsection {Breadth-First-Search Results}
Figure~\ref{fig:perf-bfs-small} shows the performance of breadth-first-search on various systems.
The semi-external system FlashGraph is generally the fastest, and GraphChi is generally the slowest.
X-Stream's performance suffers as the active list becomes sparse, because it traverses the entire graph at every iteration.

\subsubsection{Betweenness-Centrality Results}
Figure~\ref{fig:perf-bc-small} shows the performance of betweenness-centrality on various systems.
Similar to the other algorithms, FlashGraph shows the fastest performance, while X-Stream suffered from the sparse active list.
X-Stream also ran out of storage space during execution of Kron30 due to intermediate data it generated, and failed to finish.

\subsubsection {Further Discussion of PageRank}

\begin{figure}[tb]
	\begin{center}
	\includegraphics[width=0.4\textwidth]{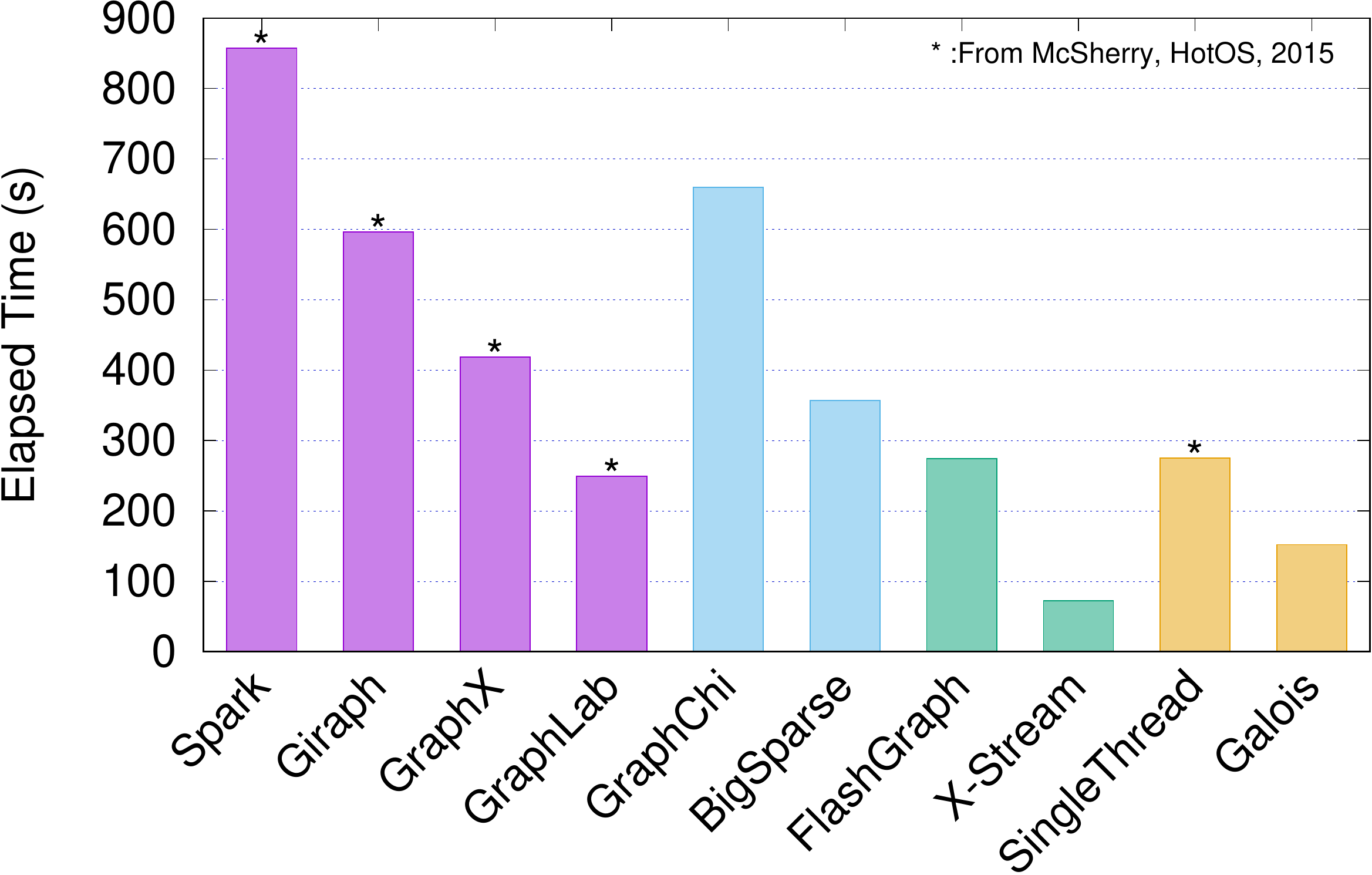}
	\caption{Execution time of PageRank on the twitter graph until convergence}
	\label{fig:compare-pagerank-small}
	\vspace{-10pt}
	\end{center}
\end{figure}

We also compared the execution of a full execution of PageRank, because some systems like Galois do not operate in terms of supersteps, and instead generates work asynchronously.
We compared the performance of some systems executed on our experimental setup, against some systems executed on 8 nodes of Amazon's AWS cloud servers.
Systems were either run until convergence, or failing that, executed for 20 iterations.
The results can be seen in Figure~\ref{fig:compare-pagerank-small}.
The results that are not our own measurements are marked with a *.
The distributed system performance for Spark, Giraph, GraphX and GraphLab are from a 8-node cluster with a total of 128 cores and are taken from McSherry, 2015~\cite{mcsherry2015scalability}.
Fully external systems GraphChi and BigSparse, as well as semi-external systems FlashGraph and X-Stream were run on our experimentation platform.
In-memory system Galois was also run on the 8-node cluster with 128 cores.
The \emph{SingleThread} system shows the performance of a single-threaded in-memory implementation running on a high-end laptop, and is also taken from~\cite{mcsherry2015scalability}.

Some of the distributed systems listed performed poorly compared to non-distributed systems, due to their distribution overhead.
Such systems provide high scalability so that bigger problems can be dealt with by providing larger clusters, but their distribution overhead harms their performance on such small clusters.
Generally for small graphs, semi-external systems perform faster than fully external ones, and in-memory systems perform better than semi-external ones.
X-Stream performs very fast, because with such a small dataset, it can optimize its cache usage and effectively become a single-node in-memory system without the distribution overhead.

We must be aware that both due to different machine configurations and organization of computation, the numbers shown in Figure~\ref{fig:compare-pagerank-small} cannot be simply taken at face value.
Instead, it should only be taken as an estimate of how a simple deployment of each systems perform.

\begin{figure*}[tb]
	\centering
	\subfloat[PageRank iteration execution time on machines with various amounts of memory] {
		\includegraphics[width=.45\textwidth]{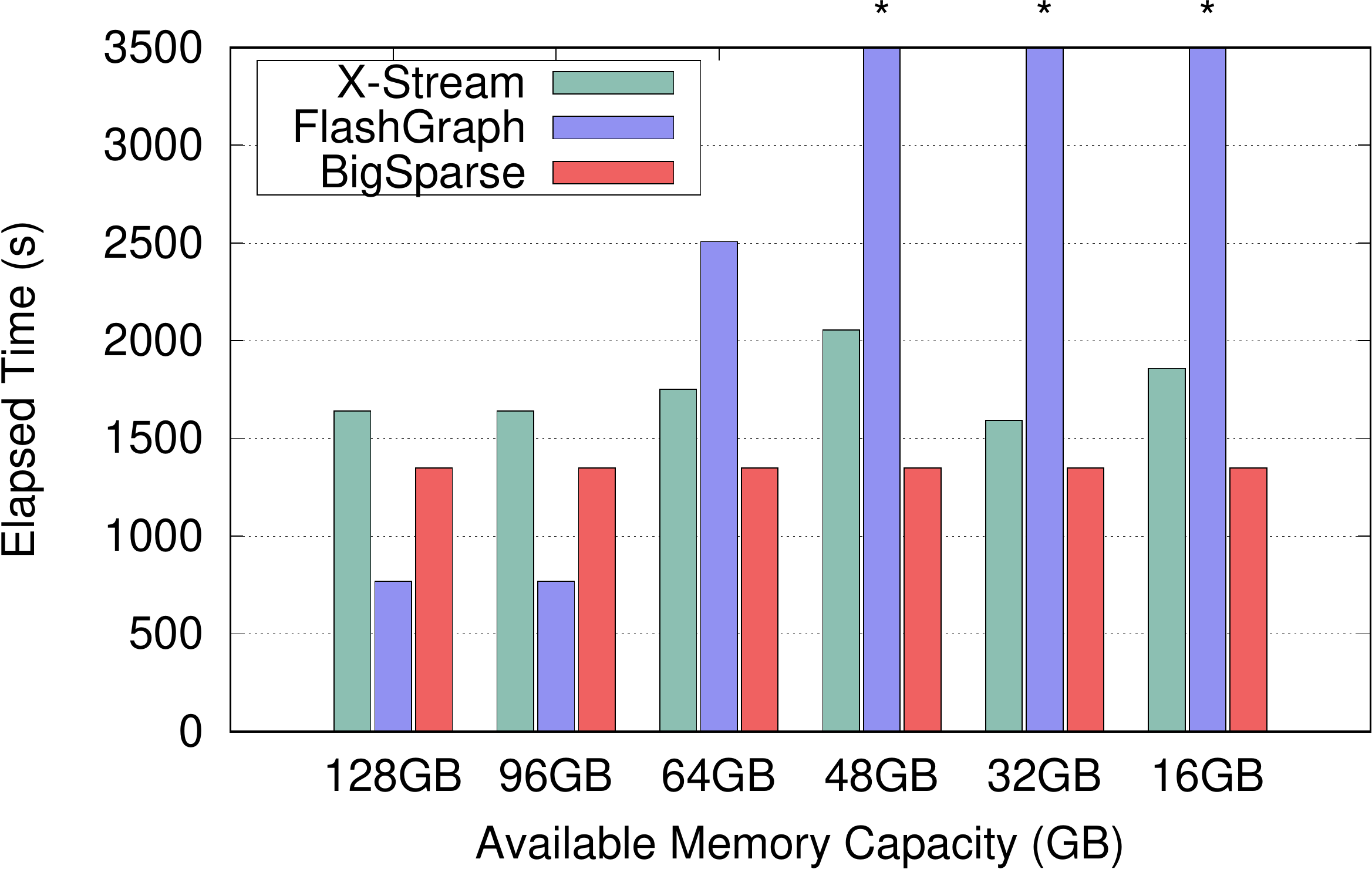}
		\label{fig:perf-pagerank}
		\vspace{-10pt}
	}
	\hfill
	\subfloat[Breadth-First-Search execution time on machines with various amounts of memory] {
		\includegraphics[width=.45\textwidth]{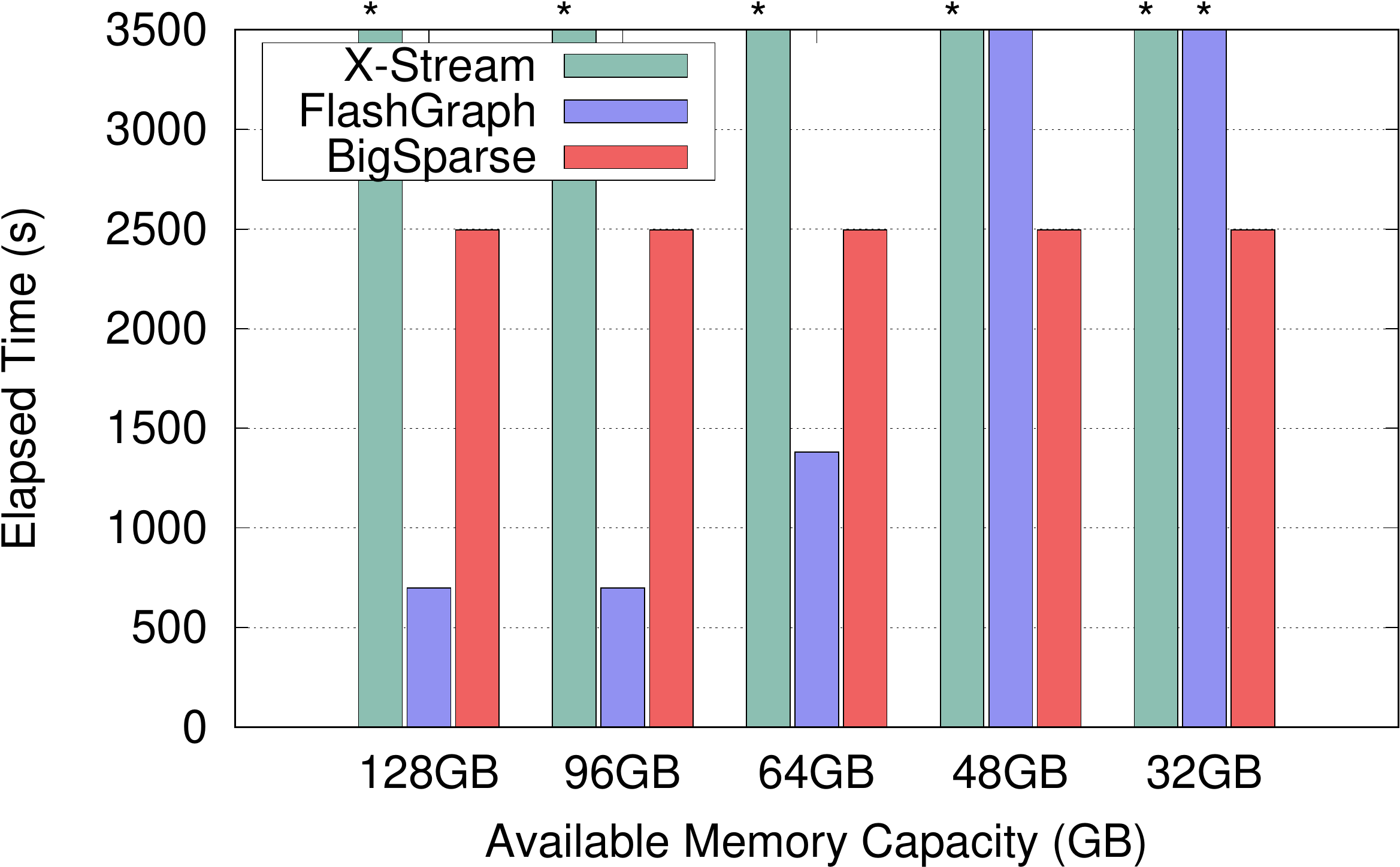}
		\label{fig:perf-bfs}
		\vspace{-10pt}
	}
	\hfill
	\subfloat[Betweenness-Centrality execution time on machines with various amounts of memory] {
		\includegraphics[width=.45\textwidth]{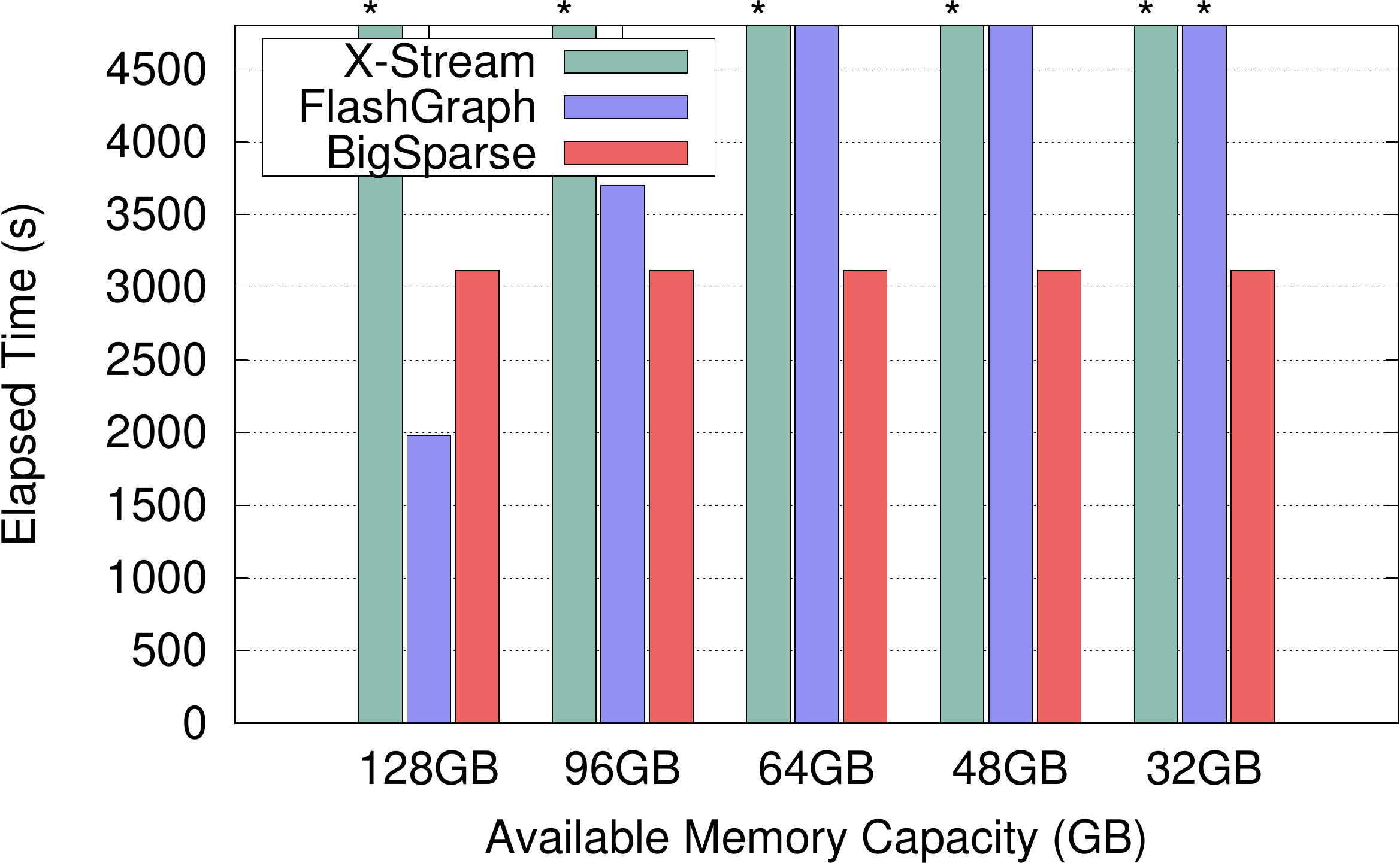}
		\label{fig:perf-bc}
		\vspace{-10pt}
	}
	\hfill
	\subfloat[Performance of the algorithms on a machine with 64GB of memory] {
		\includegraphics[width=.45\textwidth]{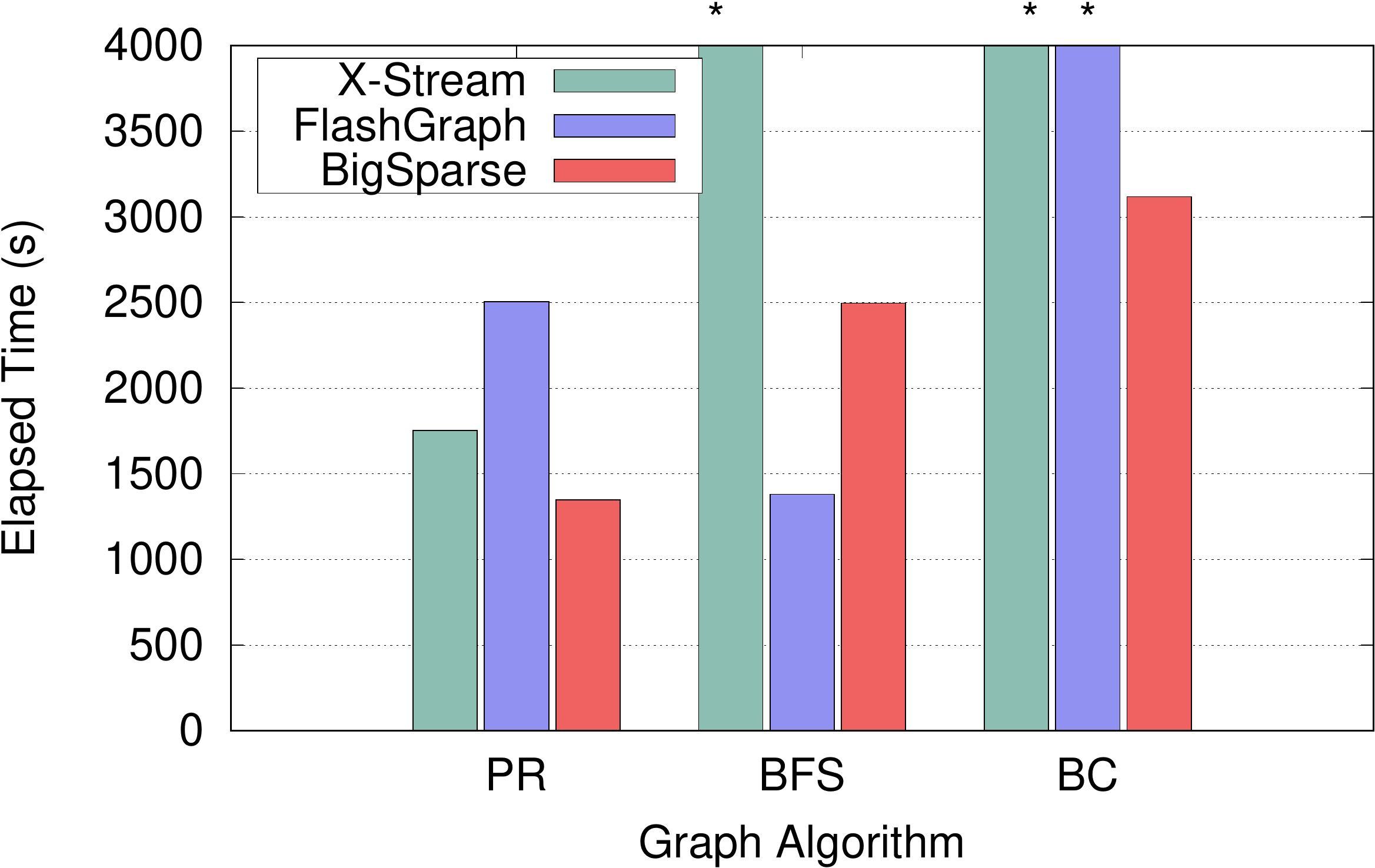}
		\label{fig:perf-64}
		\vspace{-10pt}
	}
	\hfill 

	\caption{Execution time of graph algorithms on the Web Data Commons graph (Lower is faster)}
	\vspace{-10pt}
	\label{fig:perf-billion}
\end{figure*}

\subsection {Evaluation with Graph with Billions of Vertices}
We were not able to measure the performance of GraphChi for such large graphs because we could not provide enough storage for GraphChi to import the dataset.
However, it is not difficult to make an educated guess of its performance from the small graph examples above.
We measured the performance of the remaining three systems using the three graph algorithms introduced above.

\subsubsection{PageRank Results}
The performance of the three systems running PageRank can be seen in Figure~\ref{fig:perf-pagerank}.
With enough memory capacity, FlashGraph shows the fastest performance, and X-Stream shows the slowest performance.
However, as the machine's memory capacity decreases, FlashGraph's performance degrades sharply.
The performance crossover happens on the 64GB machine, and we were unable to measure the exact execution time of FlashGraph in a reasonable amount of time.
The execution instances marked with * were were stopped manually because it was taking too long.
X-Stream partitions the graph into two and four partitions in 32GB and 16GB machines, respectively, in order to fit each partition in available memory, and shows consistent performance.
Because X-Stream logs the vertex updates between partitions, we had to add additional storage to accommodate the logs.
BigSparse also shows consistent performance regardless of available memory.



\subsubsection{Breadth-First-Search Results}
Figure~\ref{fig:perf-bfs} shows the performance of breadth-first-search of FlashGraph and BigSparse on various system configurations.
Because the active vertex list during BFS execution is usually small relative to the graph size, the memory requirements of a semi-external system like FlashGraph is also relatively low.
As a result, FlashGraph demonstrates the fastest performance even on machines with relatively small memory capacity.
The performance crossover happens on the machine with 48GBs of memory, after which the system did not finish execution in a reasonable amount of time.
Again, the execution instances marked with * were were stopped manually because it was taking too long.

We were unable to measure the performance of X-Stream for the BFS algorithm in a reasonable amount of time for any configuration.
Our observations showed execution of BFS on the WDC graph has a very long tail, where there were thousands of iterations with only a handful of active vertices.
Because X-Stream's edge-centric execution model iterates through the entire edge list at every algorithm cycle, each of the very small algorithm iterations required a non-negligible amount of execution time.
We observed each iteration taking about 500 seconds, which will result in the execution time exceeding two million seconds, or 23 days.

\subsubsection{Betweenness-Centrality Results}
Figure~\ref{fig:perf-bc} shows the performance of betweenness-centrality with FlashGraph and BigSparse.
Similarly to the other algorithms, FlashGraph shows the best performance when there is enough memory.
However, the memory requirements of BC is high, resulting in an early performance crossover, on the 96GB machine.
The high memory requirement is most likely because BC requires an additional step where each active vertex list to be transformed and sorted for back propagation.
We were unable to measure the performance of X-Stream for BC for the same reason with BFS.


Figure~\ref{fig:perf-64} shows the performance comparisons of all systems on a reasonably affordable machine with 64GBs of memory.
Because of the memory requirements of semi-external systems, BigSparse outperforms all systems except on BFS, where the active vertex list is usually small.


\subsubsection{Discussion of System Resource Utilization}

Table~\ref{tab:resourceutil} shows the systems resource utilizations of the three systems mentioned above, while running at full capacity PageRank on the WDC graph.
Both FlashGraph and X-Stream attempted to use all of the available 32 cores' CPU resources.
It is most likely because these systems spawn a lot more threads than BigSparse. As a result, both systems record 3200\% CPU usage while running.
BigSparse tries to use all of the available processing resources, but does exceed 1800\% because performance is usually bound by storage performance.

\begin{table}[h]
\centering
\begin{tabular}{|c||r|r|r|r|r}
\hline
name & Memory & Flash Bandwidth & CPU \\
\hline
BigSparse & 8GB & 500MB/s$^*$, 4GB/s$^{**}$ & 1800\% \\
FlashGraph & 60GB & 1.5GB/s & 3200\% \\
X-Stream & 80GB & 2GB/s & 3200\% \\
\hline
\multicolumn{4}{r}{$^*$During intermediate list generation} \\
\multicolumn{4}{r}{$^{**}$During external merge-reduce} \\

\end{tabular}
\caption{System resource utilization during PageRank on WDC}
\label{tab:resourceutil}
\vspace{-10pt}
\end{table}

\subsection{Benefits of Interleaving Reduction with Sorting}

One of the big factors of BigSparse's high performance is the effectiveness of interleaving reduction operations with sorting.
Interleaving sorting and reduction can reduce data movement to and from storage by as much as 90\% over if we has first sorted, and then reduced the data.

\begin{figure}[tbp]
	\begin{center}
	\includegraphics[width=0.4\textwidth]{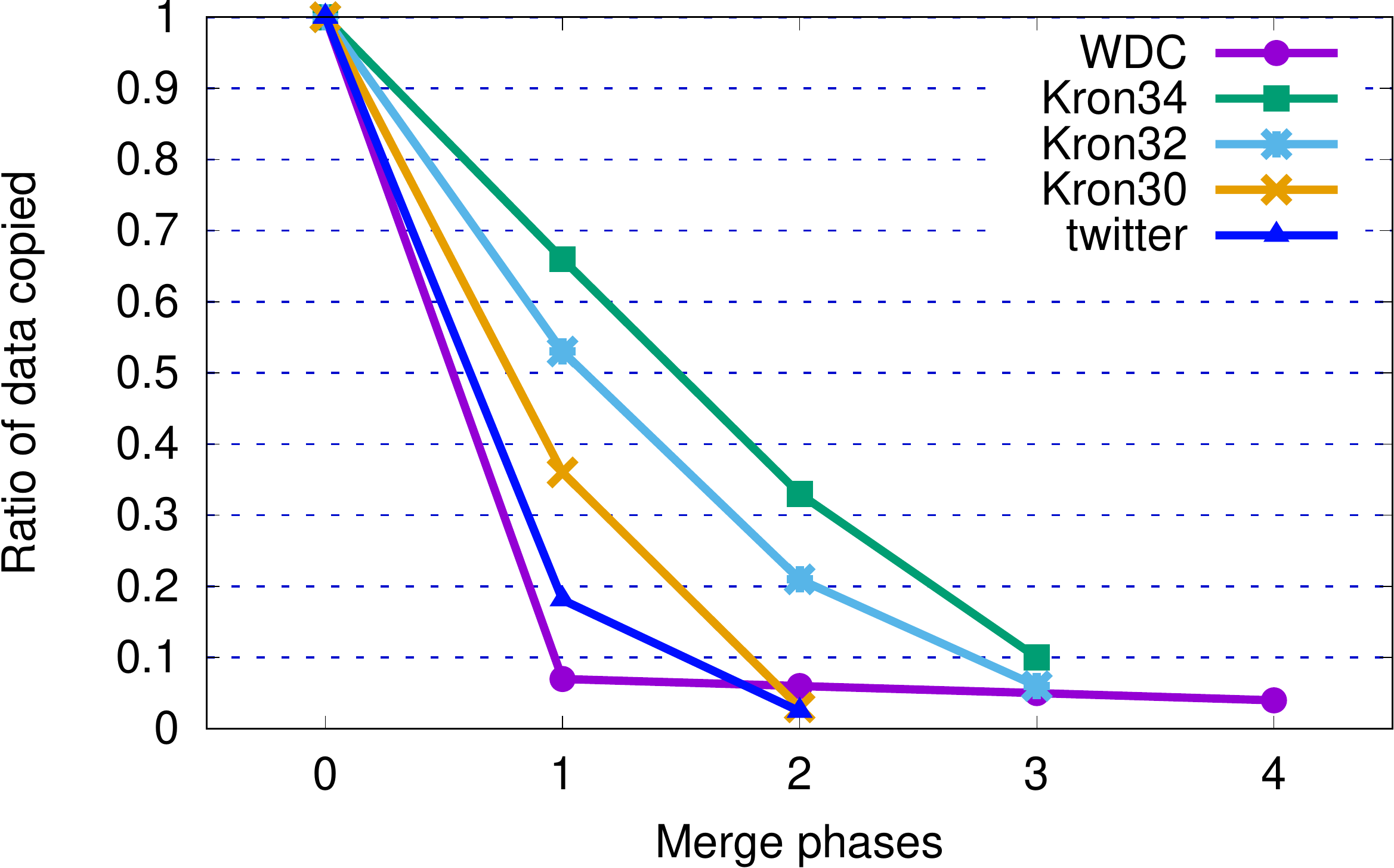}
	\caption{The fraction of data that is written to storage after each merge-reduce phase}
	\label{fig:merge-ratio}
	\vspace{-10pt}
	\end{center}
\end{figure}

Figure~\ref{fig:merge-ratio} shows the amount of data that was reduced by applying reduction to the intermediate data at every merge sort phase, starting from the intermediate list generated when all nodes are active.
As merge-reduce is progressively applied, the amount of data that is written back to storage is decreased significantly at every merge phase until the entire intermediate list is completely merged and reduced.
The number of merge-reduce phases required until the intermediate data is completely merged depends on the size of the graph.

The reduction is especially significant in the case of the two real world graphs, the twitter graph and WDC graph, where the size of the intermediate list has already been reduced by over 80\% and 90\% after the very first, in-memory merge phase.
Comparing the total amount of data that has to be written, the amount of has been reduced to 10\% and 5\% for the twitter graph and WDC, respectively.
This level of quick reduction is the result of high locality in the graph, which is a characteristic of some important real-world graphs.
In total, the overhead of moving data has been reduced to a fraction of what would have been required, if the whole list had been sorted up front.

It can be seen that interleaving reduction operations to sorting has significant benefits.
Simply comparing the amount of data that must be read and written, especially in a storage bottlenecked system, reducing the total size of data movement to 10\% is an immense benefit, and minimizes the performance impact of sorting.
An additional benefit to the reduction in data size is extending flash storage lifetime.
Because the very first intermediate data write to flash happens after the in-memory merge, after which the data size have been reduced to less than 10\%, the actual amount writes that is done to flash is low.

\section{Conclusion and Future Work}
\label{sec:conclusion}

In this paper, we have presented BigSparse, a fully external graph analytics system that can perform high-speed analytics on graphs with billions of vertices, on an affordable machine with very small memory, that is, a single Xeon server with 16GB of DRAM, and 5 SSDs.
We have shown that the performance of vertex-centric semi-external systems that support selective edge data access like FlashGraph degrade sharply when available memory is too small to accommodate vertex data.
Systems that do not support selective edge access, like X-Stream and GraphChi can maintain performance with small amount of memory using partitioning, but are inefficient for algorithms with sparse active vertices.
We have shown that BigSparse is able to pick up where semi-external systems like FlashGraph and X-Stream left off, and continue to provide high performance when the vertex data size is too large for semi-external systems. 
Unlike X-Stream and GraphChi, BigSparse provides a flexible environment that supports selective edge accesses. 
The impact of BigSparse is that it may allow, for example, bioinformatics applications to run on small machines, which may fit in a doctor's office.

We think that BigSparse can easily be scaled horizontally to run on a cluster, by partitioning the intermediate update list across machines.
The BigSparse idea is also directly realizable in the form of hardware accelerators, which may eliminate the need for a server, and drastically reduce the power consumption.
BigSparse should also work well on GPUs, which can accelerate the sorting and merging operations to make better use of flash bandwidth.

%
%
%
%
%

\vfill
\balance

\bibliographystyle{abbrv}
\bibliography{references}  

%
%
%
%
%
%

\end{document}